\documentclass[journal,onecolumn]{IEEEtran}
\usepackage{graphicx}
\usepackage{subfigure}
\usepackage{float}
\usepackage{amsmath}
\usepackage{epstopdf}
\usepackage{cases}
\usepackage{color}
\usepackage{algorithm}
\usepackage{algpseudocode}
\usepackage{verbatim}   
\usepackage{cite}
\ifCLASSINFOpdf
\else
\fi
\usepackage{caption}
\usepackage{mathtools}

\begin{document}

%
\title{Collaborative Precoding Design for Adjacent \\ Integrated Sensing and Communication \\ Base Stations}
%
%
%

\author{
	Wangjun Jiang,~\IEEEmembership{Student Member,~IEEE,}
	Zhiqing Wei,~\IEEEmembership{ Member,~IEEE,} \\
	Fan Liu,~\IEEEmembership{ Member,~IEEE,}
	Zhiyong Feng,~\IEEEmembership{Senior Member,~IEEE,}
	Ping Zhang,~\IEEEmembership{Fellow,~IEEE,}
	\\
	\thanks{This work was supported by the National Key Research and Development Program under Grant 2020YFA0711302, and the BUPT Excellent Ph.D. Students Foundation under Grant CX2022207. \emph{Corresponding author: Zhiyong Feng and Zhiqing Wei.}
		\\
  A conference version of this paper has appeared in the 2022 IEEE 95th Vehicular Technology Conference (VTC2022-Spring) \cite{[VTC]}. \\
	Wangjun Jiang, Zhiqing Wei and Zhiyong Feng are with Beijing University of Posts and Telecommunications, Key Laboratory of Universal Wireless
		Communications, Ministry of Education, Beijing 100876, China (Email: \{jiangwangjun, weizhiqing, fengzy\}@bupt.edu.cn). \\
		Fan Liu is with the Department of Electrical and Electronic Engineering,
		Southern University of Science and Technology, Shenzhen 518055, China (Email: liuf6@sustech.edu.cn). \\
		Ping Zhang is with the School of Information and Communication Engineering, Beijing University of Posts and Telecommunications, and
		State Key Laboratory of Networking and Switching Technology, Beijing 100876, China (Email: pzhang@bupt.edu.cn).
		}

}

\maketitle

\begin{abstract}
\fontsize{9pt}{6.5pt} Integrated sensing and communication (ISAC) base stations can provide communication and wide range sensing information for vehicles via downlink (DL) transmission, thus enhancing vehicle driving safety.
One major challenge for realizing high performance communication and sensing is how to deal with the DL mutual interference among adjacent ISAC base stations, which includes not only communication related interference, but also radar sensing related interference.
In this paper, we establish a DL mutual interference model of adjacent ISAC base stations, and analyze the relationship for mutual interference channels between communications and radar sensing.
To improve the sensing and communication performance, we propose a collaborative precoding design for coordinated adjacent base stations to mitigate the mutual interference under the transmit power constraint and constant modulus constraint, which is formulated as a non-convex optimization problem.
We first relax the problem into a convex programming by omitting the rank constraint, and propose a joint optimization algorithm to solve the problem.
We furthermore propose a sequential optimization algorithm, which divides the collaborative precoding design problem into four subproblems and finds the optimum via a gradient descent algorithm.
Finally, we evaluate the collaborative precoding design algorithms by considering sensing and communication performance via numerical results.
\end{abstract}


%
\IEEEpeerreviewmaketitle

\section{Introduction}
%
%
%
%

 \subsection{Background and Motivation}

 The Tri-Level study of the causes of traffic accidents has shown that 90-93\% of vehicle incidents are caused by human errors \cite{[Driverless_Future]}, which brings forward an impending need for ensuring the driving safety. To that end, most of the current intelligent vehicles are equipped with high-precision on-board sensors.
 However, on-board sensors have shortcomings in terms of sensing range and blind spot. Recognizing this fact,
 J. Andrews {\it {et al.}} proposed the perceptive mobile network (PMN) using the integrated sensing and communication (ISAC) technology \cite{[PMC]}. ISAC base stations in PMN are installed at high points on the roadside, which can provide wide range sensing information for vehicles. Moreover, IMT-2030 (6G) working group also put forward a technical scheme to realize high performance sensing of adjacent ISAC base stations \cite{[IMT_2030]}, which enables sensing while communicating by transmitting ISAC signals via the traditional base station.

 With the development of the fifth-generation (5G) Generation mobile networks, in order to pursue high spectral efficiency, the system adopts frequency reuse networking \cite{[CoMP_1]}.
%
 High spectrum utilization enables ISAC base stations to obtain large bandwidth resources and thus better communication rates and sensing resolution performance \cite{[PMC], [PMC_2]}.
 But this method leads to the mutual interference among communication base stations. Solving the mutual interference problem among adjacent ISAC base stations is a prerequisite for achieving favorable sensing and communication performance.
 In this paper, we focus on the downlink (DL) sensing and communication in the adjacent ISAC base stations system.


 The main methods to solve the problem of mutual interference include resource allocation \cite{[Radar_1]}, interference cancellation algorithm and precoding design \cite{[Pre-yuanwei],[Pre-jiayi],[Pre-feifei]}. The resource allocation mainly eliminates interference by assigning different time, frequency, space and code resources to different base stations and users \cite{[Radar_1]}. However, this method realizes interference cancellation at the cost of reduced resource efficiency \cite{[RA_1],[RA_2]}.
 The interference cancellation algorithm mitigates interference by separating the received mixed signal directly and extracts the useful signal. However, this method requires that the interfering signal differs significantly from the useful signal in some respect.
 In this paper, we focus on studying the collaborative precoding design for adjacent ISAC base stations to solve the mutual interference problem.
 {\color{black}
 There are two main challenges of the collaborative precoding design for adjacent ISAC base stations:
 \begin{itemize}
 	\item How to establish a mutual interference model among ISAC base stations: Compared with communication-only networks, the mutual interference among ISAC base stations is more complex, including communication related interference, radar sensing related interference and the interference between sensing and communication.
 	\item How to solve the collaborative precoding design problem for cancelling the mutual interference. The collaborative precoding design problem consists of some non-convex constraints, such as rank-1 constraint.
 	Solving the non-convex optimization problem with low complexity is a key challenge in achieving collaborative precoding.
 \end{itemize}
}

  \subsection{Related Work}

 Existing studies related to the construction of mutual interference models and precoding designs can be divided into three main categories, communication-related interference, radar sensing-related interference, and ISAC interference.

 \subsubsection{Communication Related Interference}

 Communication related interference includes multiuser interference, and inter-base station communication interference.
 According to \cite{[CI_26]}, the multiuser interference that makes the received symbols deviate from the original decision threshold is called destructive interference (DI), and the interference that does not make the received symbols deviate from the original decision threshold is called constructive interference (CI).
 M. Schubert {\it {et al.}} proposed the precoding design for cancelling the multiuser CI in \cite{[SDP_37]}. Choi {\it {et al.}} formulated a CI optimization problem that minimizes a sum of minimum mean square error (MMSE) for all users, while ensuring the minimum required CI gain with a fixed total power constraint \cite{[CI_2019]}.
 In terms of cancelling the inter-base station communication interference, M. Karakayali {\it {et al.}} proposed forced zero precoding and dirty paper coding algorithms in \cite{[CoMP_20],[CoMP_21]}.
 However, these algorithms require high speed synchronous information interaction among base stations and high computational complexity for the control center.
 To fully utilize the computational power of each base station and reduce the computational complexity of the control center, \cite{[CoMP_22],[CoMP_23],[CoMP_25]} proposed the distributed precoding design method for the adjacent base stations system.


 \subsubsection{Radar Sensing Related Interference}

 Similar to communication related interference, sensing related interference contains multipath interference of radar itself and interference among radars. G. Cui {\it {et al.}} first pointed out that multipath echo is an interference for radar sensing and can lead to false targets \cite{[SC_37]}. A multiple input multiple output (MIMO) radar precoding design is proposed for cancelling the multipath interference \cite{[SC_37]}.
 In terms of cancelling the interference among radars, X. Yang {\it {et al.}} proposed an adaptive interference phase cancellation algorithm assisted by large aperture antenna array \cite{[Radar_29]}. To effectively filter out the interference signals with linear/curvilinear distribution in the time-frequency domain, a time-frequency filtering method based on the short-time Fourier transform is proposed in \cite{[Radar_71]}.

 \subsubsection{ISAC Interference}

 Compared with the above two types of interference, the ISAC interference contains not only interference within communication and radar, but also interference between communication and sensing \cite{[ISAC_2]}.
 The interference between the ISAC echo signal and uplink (UL) communication signal is a type of interference between communication and sensing.
 To avoid this type of interference, J. Andrews {\it {et al.}} put forward a procedure of communication and sensing in the time division duplex (TDD) mode, which can stagger the ISAC echo signal and UL communication signal in the time domain \cite{[TDD]}.
 Interference from frequency reuse MIMO radars to ISAC base stations is also one of the factors that limit the performance of ISAC. To cancel this type of interference, Liu {\it {et al.}} proposed a precoding design algorithm and obtain the globally optimal solutions to the precoding design problem with constant modulus constraint (CMC) and similarity constraint (SC) by branch-and-bound (BnB) algorithm, which outperforms the conventional successive Quadratic Constrained Quadratic Programming (QCQP) Refinement (SQR) algorithm \cite{[Liu_38],[Liu_1]}.

 In summary, there have been many studies on establishing the mutual interference model and precoding design for communication related interference cancellation or radar related interference cancellation only, but less research on precoding techniques for ISAC interference reduction.
 The recent overview paper \cite{[BF-ISAC-1]} discussed collaborative signal processing for wireless sensing, and a number of related works, e.g., \cite{[BF-ISAC-2]}, have also been surveyed in this overview paper.
 The procedure of communication and sensing in TDD proposed by J. Andrews {\it {et al.}} can stagger the ISAC echo signal and UL communication signal, but it {\color{black}cannot} cancel the DL communication and sensing interference among adjacent ISAC stations \cite{[TDD]}.
 In this paper, we will focus on the DL mutual interference model construction and collaborative precoding design for adjacent ISAC base stations.

 \subsection{Main Contributions of Our Work}

 Inspired by the above researches, in this paper, we
 establish a DL mutual interference model of adjacent ISAC base stations and propose a collaborative precoding design algorithm for coordinated adjacent base stations to mitigate the DL mutual interference.
 We focus on the DL scenario of adjacent ISAC base stations, and coordinate the mutual interference between base stations through collaborative precoding to realize the overall improvement of communication and sensing performance.
 The main contributions of the paper can be summarized as follows:

1. We establish a DL mutual interference channel model, by taking both DL communication interference and radar sensing interference into account. DL communication interference includes multiuser interference and inter-base station interference. Radar sensing interference includes multipath interference and echo interference among ISAC base stations.
Moreover, we analyze the relationship between communication and radar sensing mutual interference channels. By multiplexing communication mutual interference channel information, the overheads for radar mutual interference channel estimation can be greatly reduced.

2. we propose a collaborative precoding design algorithm for coordinated adjacent base stations to mitigate the mutual interference under the transmit power constraint and CMC. Due to the rank constraint and CMC, the collaborative precoding design problem for cancelling the mutual interference is a non-convex problem.
To relax the problem into a convex optimization problem, we first omit the rank constraint, and propose a joint optimization algorithm (JOA) to solve the problem. The approximated solution can be obtained by eigenvalue decomposition. For CMC, we transform CMC into SC by selecting the ideal reference waveform satisfying CMC.

3. To reduce the algorithm complexity, we furthermore propose a sequential optimization algorithm (SOA) to solve the collaborative precoding design problem, which divides the collaborative precoding design problem into four subproblems and obtain the optimal precoding design by gradient descent. The convergence proof for SOA is also provided in this paper. Moreover, we derive the computational complexity of the proposed algorithms. Numerical results verify the effectiveness of the SOA in reducing the complexity.
%
%
%
%
%

The remaining parts of this paper are organized as follows.
Section \ref{sec:system_model} describes the system model of adjacent ISAC base stations. Section \ref{sec:ISAC_MICM} elaborates the ISAC mutual interference channel model of the adjacent ISAC base stations system. The collaborative precoding design problem is formulated in Section \ref{sec:OPF}. Section \ref{sec:OWD} introduces the sequential and joint optimization algorithms. The trade-off of sensing and communication performance based on proposed collaborative precoding design is analyzed and simulated in section \ref{sec:Simulation}. Section \ref{sec:Conclusion} concludes the paper.

The symbols used in this paper are described as follows. Vectors and matrices are denoted by boldface small and capital letters; the transpose, complex conjugate, Hermite, inverse, and pseudo-inverse of the matrix ${\bf{A}}$ are denoted by ${{\bf{A}}^{T}}$, ${{\bf{A}}^*}$, ${{\bf{A}}^{H}}$, ${{\bf{A}}^{ - 1}}$ and ${{\bf{A}}^\dag}$, respectively; ${\rm{diag}}(\bf{x})$ is the operation that generates a diagonal matrix with the diagonal elements to be the elements of $\bf x$; $ \otimes $ is the Kronecker product operator; $ \odot $ is the Hadmard product operator.

\section{System Model}\label{sec:system_model}

\begin{figure}[ht]
	\includegraphics[scale=0.25]{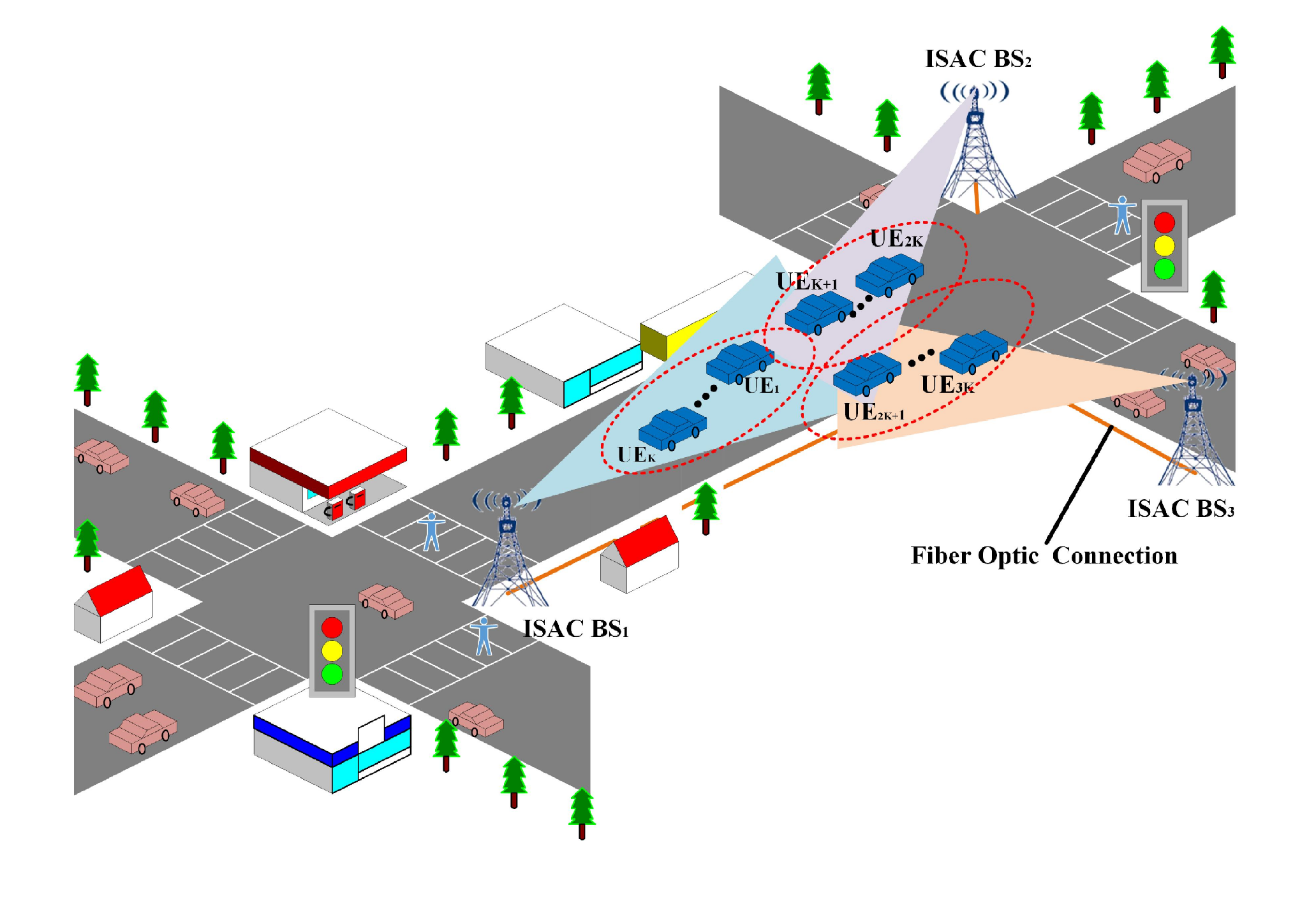}
	\centering
	\caption{Adjacent ISAC base stations system.}
	\label{fig:ISAC_system_model}
\end{figure}

\begin{figure}[ht]
	\includegraphics[scale=0.3]{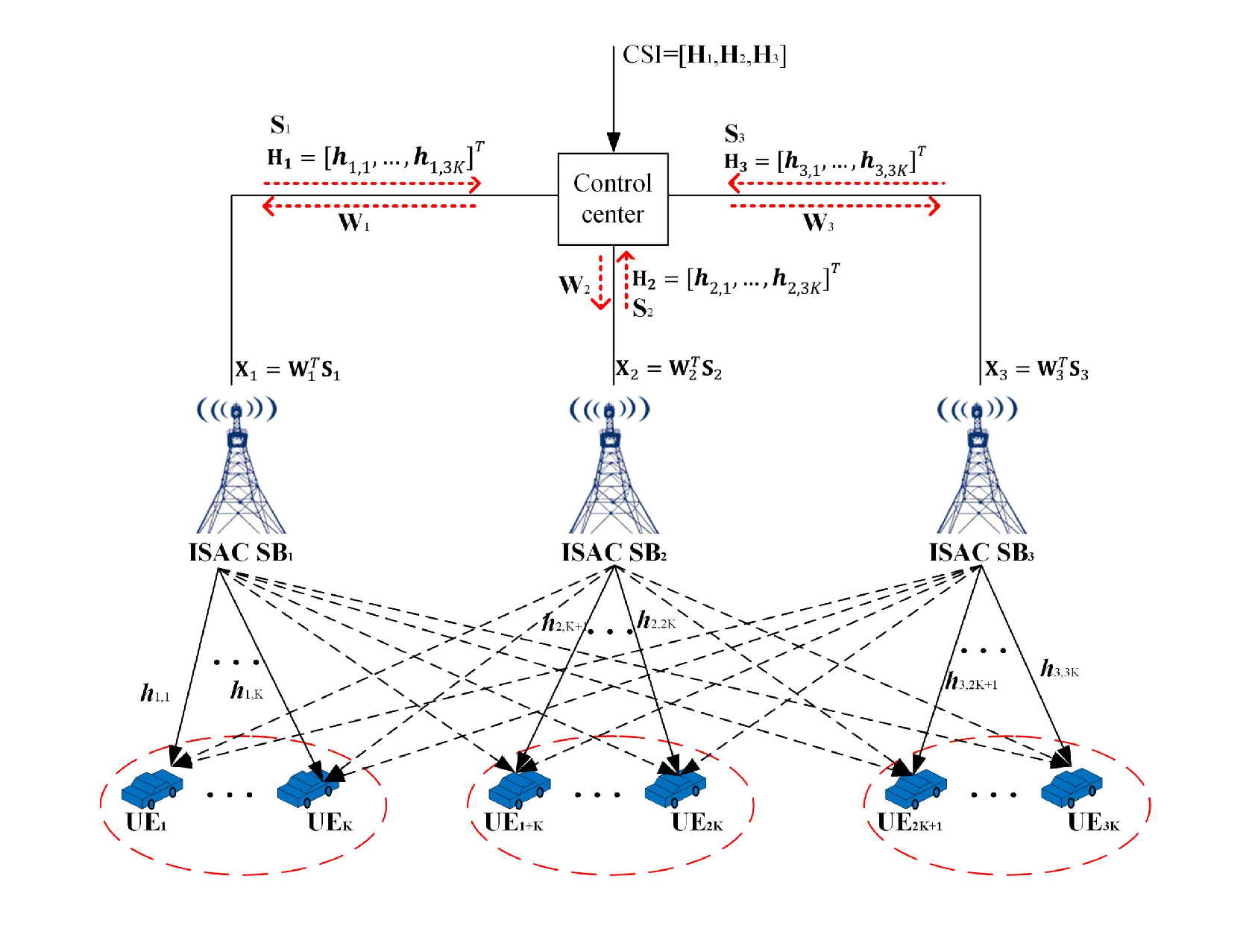}
	\centering
	\caption{Information interaction in the adjacent ISAC base stations system.}
	\label{fig:system_CSI}
\end{figure}

The adjacent ISAC base stations system is the main scenario studied in this paper. Without loss of generality, we take the scenario where three ISAC base stations provide communication and sensing services to vehicle targets as an example, as shown in Fig. \ref{fig:ISAC_system_model}.
%
In order to obtain high-performance communication and radar sensing, the base stations first send the channel estimation signal, and the receivers adopt the maximum likelihood estimation (MLE) algorithm \cite{[MLE]} to obtain the estimation of the communication and radar interference channels.
Although the target detection can be achieved based on the estimation of the radar interference channel, the performance of target estimation is poor due to the serious mutual interference among base stations.
To solve this problem, this paper proposes a collaborative precoding method for adjacent base stations to maximize the receiving signal to interference plus noise ratio (SINR) and achieve optimal communication and radar sensing.

In this paper, a mutual interference model between ISAC base stations is established, and a collaborative precoding algorithm is proposed based on the mutual interference channel parameters to obtain better communication and target detection performance.
Users can estimate the communication interference parameters by receiving the DL channel estimation signal and feed it back to the ISAC base station.
ISAC base stations can obtain the sensing mutual interference channel parameters between base stations by receiving the echo of sensing signals sent by itself and adjacent base stations.
In order to achieve better communication and target detection performance, ISAC base stations design appropriate precoding according to the communication sensing mutual interference channel parameters.

Let us assume that each ISAC base station equipped with a circular center antenna array with $N_t$ antennas, which is detecting radar targets while providing communication service for $K$ single-antenna users at the same time.
{\color{black}
These $3K$ users are not only communication terminals, but also sensing targets of ISAC base station.
As communication terminals, ISAC $\rm{BS}_1$ provides communication service for the $UE_1 \sim UK_K$, ISAC $\rm{BS}_2$ provides communication service for the $UE_{K+1} \sim UK_{2K}$, and ISAC $\rm{BS}_3$ provides communication service for the $UE_{2K+1} \sim UK_{3K}$.
As sensing targets, each ISAC base station needs to detect all $3K$ targets.
The larger the target can obtain SINR, the better the sensing performance that can be obtained. For ideal point targets, the corresponding sensing SINR greater than or equal to 13.2 dB is required to achieve a detection probability of 95\% at a false alarm probability of $10^{-5}$ \cite{[radar_SINR]}.
}
Moreover, ISAC base stations are connected to each other by fiber optics to meet the transmission requirements for synchronization and channel parameters sharing for inter-base station collaboration.
As shown in Fig. \ref{fig:system_CSI}, ISAC base stations first transmit their channel parameters $\bf H$ to the control center, which performs collaborative precoding design, generates the precoding matrix $\bf W$ of the antennas, and returns it to each ISAC base station. Then, each base station generates the transmit ISAC signal $\bf X$ by $\bf W$.
It should be noted that some radar sensing channel parameters are not need to be shared among ISAC base stations because they can be inferred from the communication channel state information, which will be described in detail in section \ref{sec:ISAC_MICM}.

\subsection{ISAC Mutual Interference Channel Model}\label{sec:ISAC_MICM}

Based on the above analysis in section \ref{sec:system_model}, ISAC base stations need to share channel parameters for collaborative precoding.
In this subsection, we analyze the relationship between communication channel parameters and radar sensing channel parameters.

\subsubsection{Communication interference channel model}\label{sec:ISAC_MICM_1}
\begin{figure}[htbp]
	\centering
	\begin{minipage}[t]{0.48\textwidth}
		\centering
		\includegraphics[width=7.5cm]{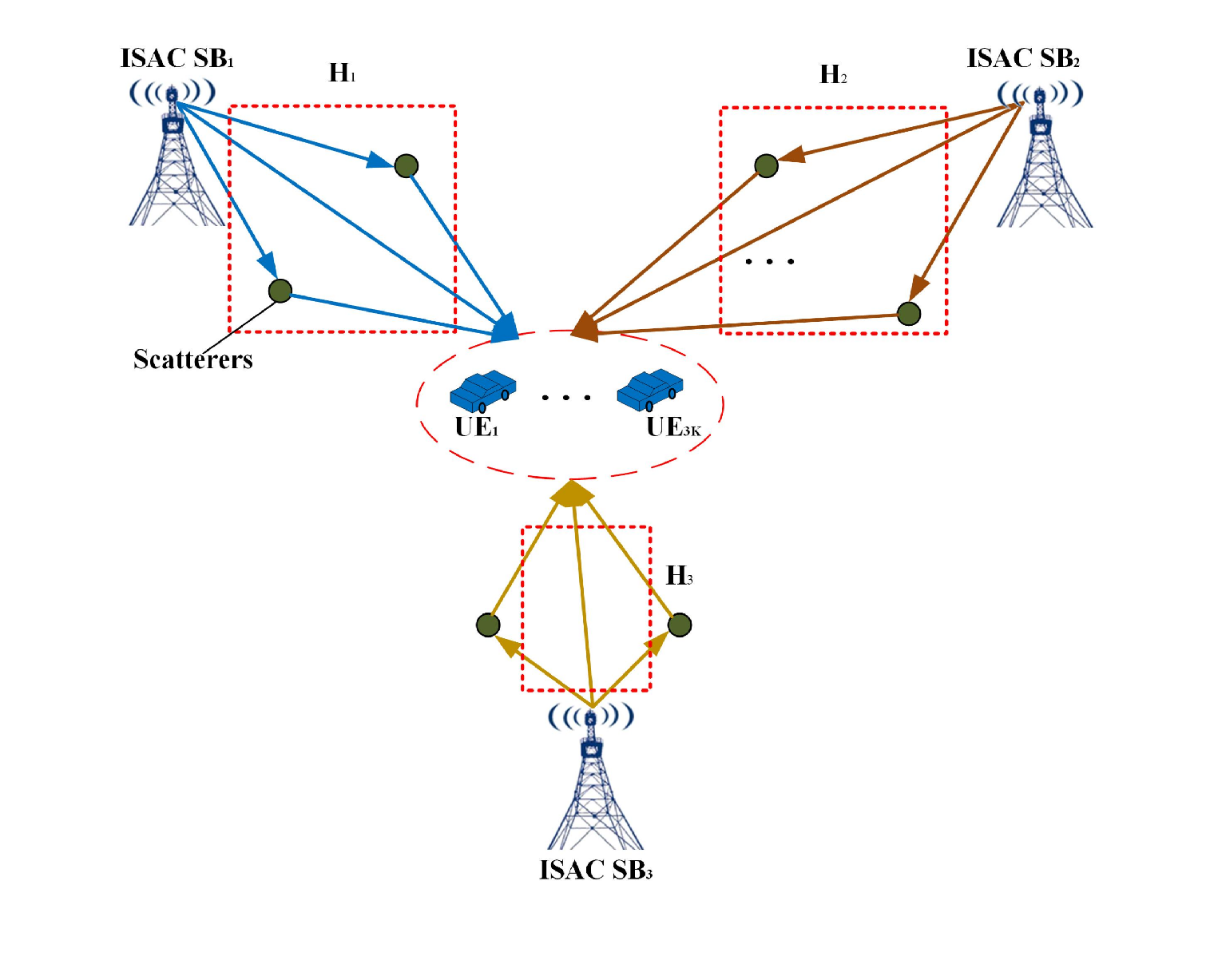}
		\caption{Communication interference channel model for users.}
		\label{fig:DL_channel}
	\end{minipage}
	\begin{minipage}[t]{0.48\textwidth}
		\centering
		\includegraphics[width=7.5cm]{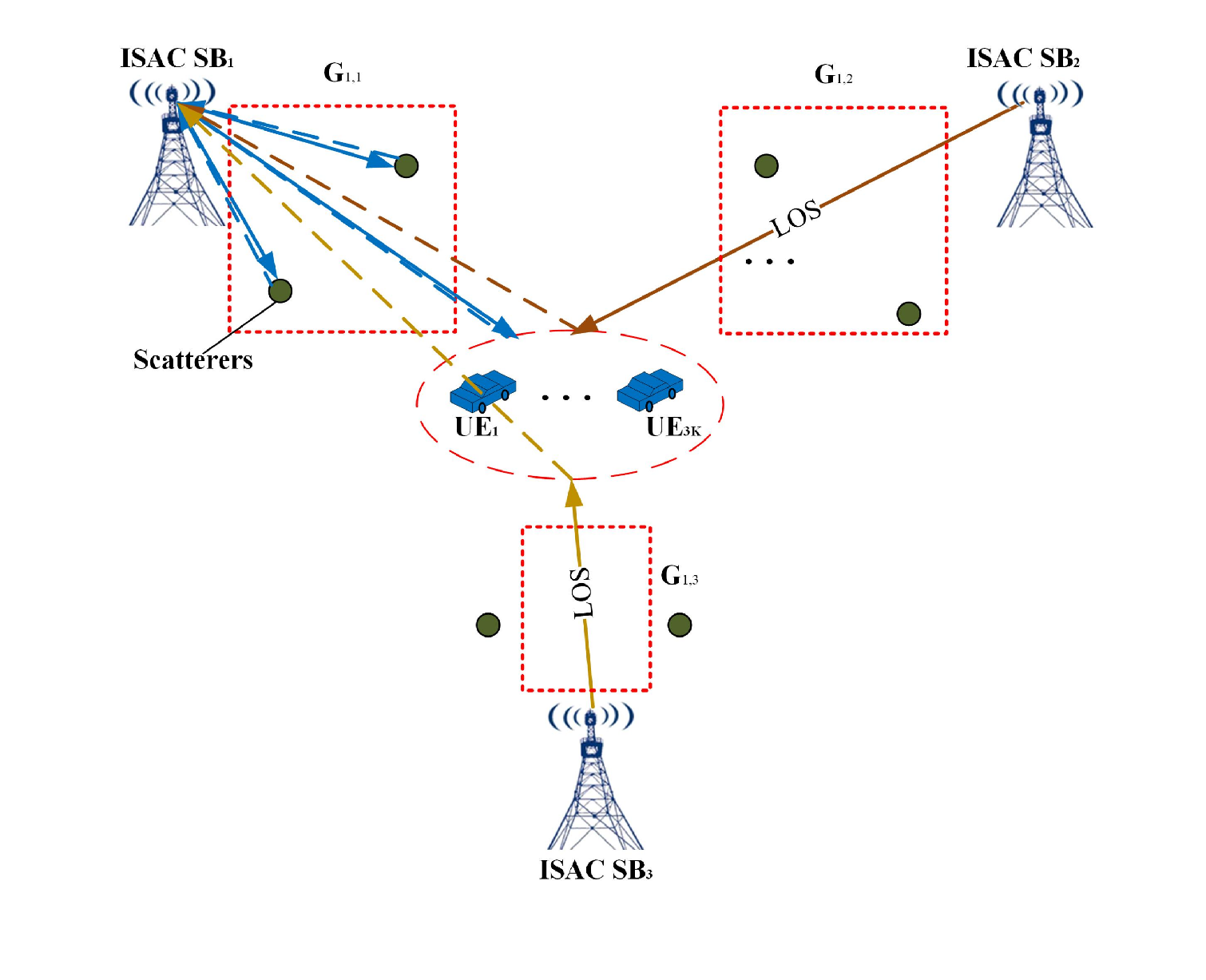}
		\caption{Radar sensing interference channel model for ISAC ${\rm{BS}}_1$.}
		\label{fig:Radar_channel}
	\end{minipage}
\end{figure}

As shown in Fig. \ref{fig:DL_channel}, communication channel parameters of ISAC $\rm{BS}_1$, ISAC $\rm{BS}_2$ and ISAC $\rm{BS}_3$ can be expressed as
\begin{equation}\label{equ:Com_CSI_1}
\begin{aligned}
{{\bf H}_{j=1,2,3}} &= [{\bf h}_{j,1},{\bf h}_{j,2},…,{\bf h}_{j,3K}], \in {{\mathcal C}^{N_t \times 3K}}
\end{aligned}.
\end{equation}
Assuming that each channel has $L_p$ propagation paths, then the DL communication sub-channel ${\bf h}_{j,i}$ from ISAC ${\rm BS}_j$ to the $i$-th user can be expressed as
\begin{equation}\label{equ:Com_CSI_2}
{\bf h}_{j,i} = \sqrt{N_t} \sum_{l=0}^{L_{p}-1} {\alpha_{j,i,l}^c {\bf a}(\theta^c_{j,i,l}, \phi^c_{j,i,l})}, \in {{\mathcal C}^{N_t \times 1}},
\end{equation}
where $\alpha_{j,i,l}^c = \widetilde{f}_{j,i,l}^c  \widetilde{F}_{j,i,l}^c e^{(-j\Delta \varphi_{j,i,l}^c)}$ denotes the total fading factor of the {\color{black}$l$-th} DL communication path, and $\Delta \varphi_{j,i,l}^c = \frac{2\pi f_c d_{c,j,i,l}}{c_0}$ is the phase offset, with $d_{c,j,i,l}$ being the length of the $l$-th communication path from the $i$-th user to ISAC ${\rm BS}_j$, $c_0$ being the speed of light and $f_c$ being the frequency of ISAC signals, $\widetilde{f}_{j,i,l}^c$ is the {\color{black}small-scale} fading factor, which obeys the Weibull distribution \cite{[Weibull_181]}.
The {\color{black}large-scale} fading factor is $\widetilde{F}_{j,i,l}^c = \frac{\lambda^2}{(4\pi)^2 {d_{c,j,i,l}^2}}$, where $\lambda$ is the wavelength of the ISAC signal.
It should be noted that the number of propagation paths of each channel may be different, but this has no impact on the research in this paper. Therefore, for the convenience of expression, this paper assumes that each channel has $L_p$ propagation paths.

\begin{figure}[ht]
	\includegraphics[scale=0.2]{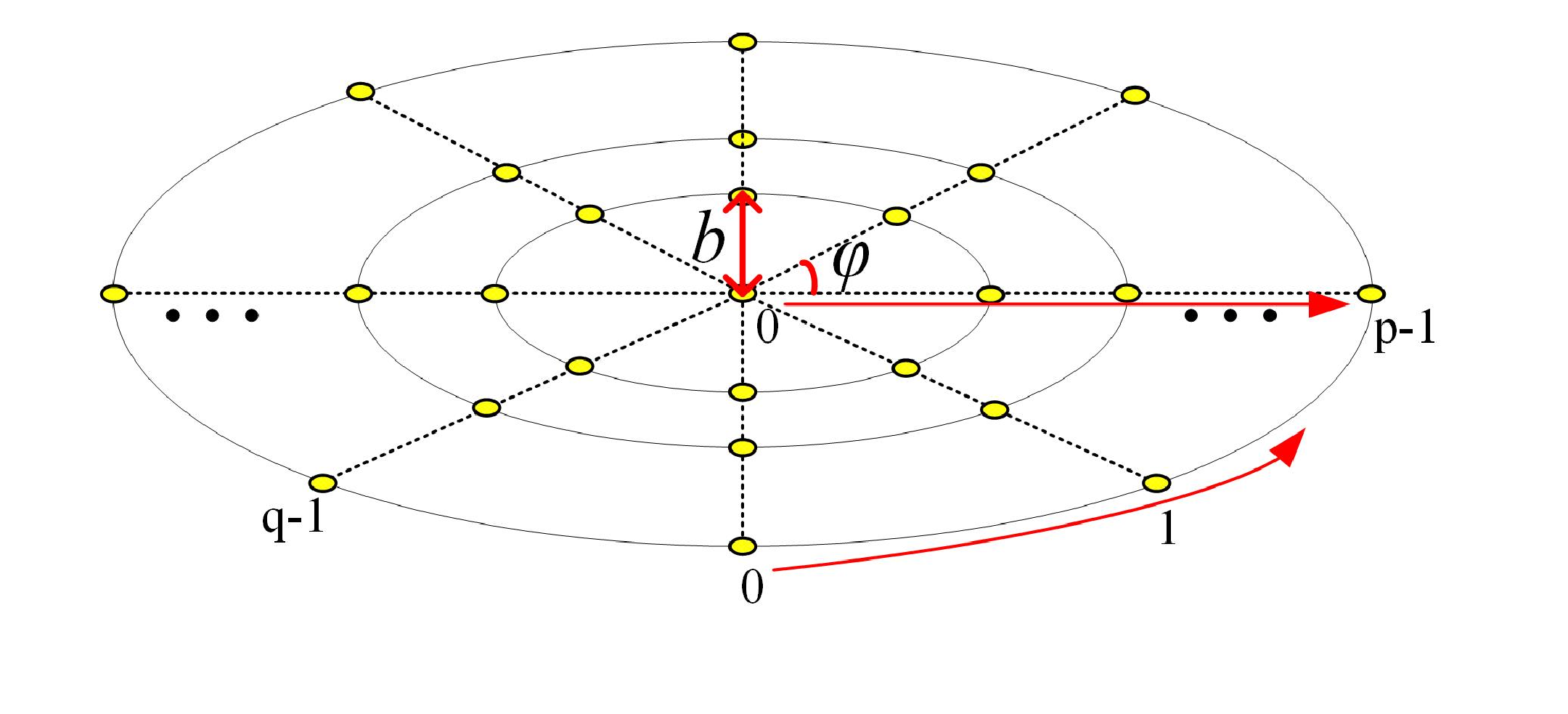}
	\centering
	\caption{Uniform circular antenna array.}
	\label{fig:system_ant}
\end{figure}
{\color{black}
To generate three-dimensional (3D) downward sensing beam, the uniform
circular antenna (UCA) is adopted in the adjacent ISAC base stations system \cite{[JWJ]}. Compared with the traditional uniform planar antenna (UPA), UCAs have two minor differences in beamforming:
\begin{itemize}
	\item compared with UPAs, the beamforming algorithm based on the UCA is a bit more complicated, because a coordinate change is required, as shown in \eqref{equ:Com_CSI_7}.
	\item Using the same number of antennas, UCAs can generate a beam with lightly smaller sidelobe than UPAs, and the related work can be referred to Fig. 5(a) in \cite{[UCA]}.
\end{itemize}
 }
As Fig. \ref{fig:system_ant} shows, there are several layers in the array. The center antenna is selected as the phase reference antenna (PFA). From the center to the periphery, the layers are labeled from $0$ to $p-1$. There are ${q}$ antenna elements in each layer, except the $\it 0$-th layer that has only one antenna element. And the number of antennas is $N_t = (p-1) q + 1$.
Then, the steering vector of the transmitting antenna is
\begin{equation}\label{equ:Com_CSI_6}
	\begin{aligned}
		{\bf a}(\theta^c_{1,i,l}, \phi^c_{1,i,l}) &= [{\rm{1}},{\it{a}_{\rm{1,0}}}({\phi^c_{1,i,l}},{\theta^c_{1,i,l}}),..., {\it{a}_{{\it{m},{\it n}}}}({\phi^c _{{1,i,l}}},{\theta^c _{{1,i,l}}})\\
			& ,..., {\it{a}_{\it{p} - \rm 1,{\it q} - 1}}({\phi^c_{{1,i,l}}},{\theta^c_{{1,i,l}}})]^{\it{T}}, \in {{\mathcal C}^{N_t \times 1}}
	\end{aligned},
\end{equation}
where ${\it{a}_{{\it{m},{\it n}}}}(\theta^c_{1,i,l}, \phi^c_{1,i,l})$ is the phase difference between the ${\it{n}}$-th antenna element of the ${\it{m}}$-th layer and PFA, can be expressed as
\begin{equation}\label{equ:Com_CSI_7}
\begin{aligned}
{\it{a}_{{\it{m},{\it n}}}}(\theta^c_{1,i,l}, \phi^c_{1,i,l}) &= {\color{black}\exp}( - j\frac{{2\pi }}{\lambda } {\bf{u}} _{m,n}^T{ {\bf{v}}_{1,i,l}}), \\
{{\bf{v}}_{1,i,l}} &= {[{\cos} {\phi^c_{1,i,l}}{\sin}{\theta^c_{1,i,l}},{\sin}{\phi^c _{1,i,l}}{\sin}{\theta^c_{1,i,l}}]^T}, \\
{\bf{u}} _{m,n} &= {[{\cos}(\frac{2\pi n}{q})mb,{\sin}(\frac{2\pi n}{q})mb]^T},
\end{aligned}
\end{equation}
where $\lambda$ is the waveform length of the ISAC signal, ${\it{b}}$ is the distance between two adjacent layers, $\theta^c_{1,i,l}$ and $\phi^c_{1,i,l}$ are the azimuth and pitch angle of the $l$-th communication path.
%

\subsubsection{Radar sensing interference channel model}\label{sec:ISAC_MICM_2}

{\color{black}
The radar sensing interference channel consists of two one-way transmission channels, each of which is similar to the communication channel. The specific channel parameters are compared in Table \ref{label:ISAC_CSI}.
}
Since radar sensing interference channel parameters of three ISAC base stations ${\bf G}_1$, ${\bf G}_2$ and ${\bf G}_3$, are expressed similarly, this subsection takes ${\bf G}_1$ as an example to analyze the radar sensing interference channel.
Since the signal attenuates greatly after reflection, we only consider the line of sight (LOS) echo channel after one reflection, as shown in Fig. \ref{fig:DL_channel} and Fig. \ref{fig:Radar_channel}.
Then, radar sensing channel parameters of ISAC $\rm{BS}_1$ can be expressed as
{\color{black}
\begin{equation}\label{equ:Rad_CSI_1}
	\begin{aligned}
		{{\bf G}{_1}} &= {{\bf G}{_{1,1}}} + {{\bf G}{_{1,2}}} + {{\bf G}{_{1,3}}}, \in {{\mathcal C}^{N_t \times N_r}} \\
		{{\bf G}{_{1,j=1,2,3}}} &= \sum_{i=1}^{3K} {\bf g}_{1,j,i}
	\end{aligned},
\end{equation}
where $N_t$ and $N_r$ are the number of received and transmitted antenna array elements. Without loss of generality, assume that $N_t = N_r$.
The radar sensing sub-channel of the $i$-th target can be expressed as
\begin{equation}\label{equ:Rad_CSI_2}
	\begin{aligned}
		{\bf g}_{1,1,i} &= {\bf g}_{1,1,i,0} + \sum_{l=1}^{L_{p}-1}{\bf g}_{1,1,i,l}  \\
		&= \sqrt{N_t} \sigma_{i,l} {\alpha_{1,i,0}^r {\bf a}(\theta^r_{1,i,0}, \phi^r_{1,i,0})  {\bf a}(\theta^r_{1,i,0}, \phi^r_{1,i,0}) } \\
		& + \sqrt{N_t}\sum_{l=1}^{L_{p}-1} \sigma_{i,l} {\alpha_{1,i,l}^r {\bf a}(\theta^r_{1,i,l}, \phi^r_{1,i,l})  {\bf a}(\theta^r_{1,i,l}, \phi^r_{1,i,l}) }  \\
		{\bf g}_{1,2,i} &= \sqrt{N_t} \sigma_{i,0} {\alpha_{2,i,0}^r {\bf a}(\theta^r_{1,i,0}, \phi^r_{1,i,0})  {\bf a}(\theta^r_{1,i,0}, \phi^r_{1,i,0}) }  \\
		{\bf g}_{1,3,i} &= \sqrt{N_t} \sigma_{i,0} {\alpha_{3,i,0}^r {\bf a}(\theta^r_{1,i,0}, \phi^r_{1,i,0}) {\bf a}(\theta^r_{1,i,0}, \phi^r_{1,i,0}) }
	\end{aligned},
\end{equation}}
where $\sigma_{i,l} $ is the reflection cross section (RCS) of the $i$-th target from the $l$-th path, $\alpha_{1,i,l}^r = \widetilde{f}_{1,i,l}^r  \widetilde{F}_{1,i,l}^r e^{(-j\Delta \varphi_{1,i,l}^r)}$ denotes the total fading factor of the $l$-th echo path,
where $\Delta \varphi_{1,i,l}^r = \frac{2\pi f_c 2 d_{r,1,i,l}}{c_0}$ is the phase offset, $\widetilde{F}_{1,i,l}^r = \frac{\lambda^2}{(4\pi)^3 {d_{r,1,i,l}^4}}$ is the {\color{black}large-scale} fading factor, with $d_{r,1,i,l}$ being the length of the $l$-th radar sensing path from the $i$-th target to ISAC ${\rm BS}_1$, $\widetilde{f}_{1,i,l}^r$ is the small-scale fading factor, which obeys the Weibull distribution.
Similarly, the total fading factor of the LOS from ISAC ${\rm BS}_2$ and ISAC ${\rm BS}_3$ can be expressed as
\begin{equation}\label{equ:Rad_CSI_4}
\begin{aligned}
\alpha_{2,i,0}^r &= \widetilde{f}_{2,i,0}^r  \widetilde{F}_{2,i,0}^r e^{(-j\Delta \varphi_{2,i,0}^r)} \\
\alpha_{3,i,0}^r &= \widetilde{f}_{3,i,0}^r  \widetilde{F}_{3,i,0}^r e^{(-j\Delta \varphi_{3,i,0}^r)}
\end{aligned},
\end{equation}
where
\begin{equation}\label{equ:Rad_CSI_5}
\begin{aligned}
\Delta \varphi_{2,i,0}^r &= \frac{2\pi f_c (d_{r,1,i,0} + d_{r,2,i,0})}{c_0} \\
\Delta \varphi_{3,i,0}^r &= \frac{2\pi f_c (d_{r,1,i,0} + d_{r,3,i,0})}{c_0} \\
\widetilde{F}_{2,i,l}^r &= \frac{\lambda^2}{(4\pi)^3 {(d^2_{r,1,i,0} d^2_{r,2,i,0})}} \\
\widetilde{F}_{3,i,l}^r &= \frac{\lambda^2}{(4\pi)^3 {(d^2_{r,1,i,0}d^2_{r,3,i,0})}} \\
\end{aligned},
\end{equation}
with $d_{r,j,i,l}$ being the length of the $l$-th radar sensing path from the $i$-th target to ISAC ${\rm BS}_j$.

Based on the above analysis, we can find that there is a great similarity between communication channel parameters and radar sensing channel parameters. By multiplexing communication mutual interference channel information, the cost of radar mutual interference channel estimation can be greatly reduced.
Take ISAC ${\rm BS}_1$ for example, the key parameters of ${\bf H}_1$ and ${\bf G}_1$ are listed in Table \ref{label:ISAC_CSI}.
The targets and scatters are in the opposite direction of the communication and radar sensing channel.
The fading factor of the communication channel is similar to that of the radar sensing channel.
The phase offset of the radar sensing channel is nearly twice that of the communication channel.
The {\color{black}small-scale} fading factors of the communication and radar sensing channel are subject to Weibull distribution.
The communication {\color{black}large-scale} fading factor obeys the law of inverse squares of the target distance, while the radar {\color{black}large-scale} fading factor is inversely proportional to the 4th power of the distance.
\begin{table}[ht]
	\caption{Relationship between communication and radar sensing channel parameters}
    \centering
	\label{label:ISAC_CSI}
	\begin{tabular}{l|l|l}
		\hline
		& Communication channel & Radar sensing channel  \\ \hline
	Angle of targets & $\theta^c_{1,i,0}$, $\phi^c_{1,i,0}$  & \begin{tabular}[c]{@{}l@{}} $\theta^r_{1,i,0} = \theta^c_{1,i,0} + \pi$ \\  $\phi^r_{1,i,0} = \phi^c_{1,i,0} + \pi$ \end{tabular}  \\ \hline
	Angle of scatterers	& $\theta^c_{1,i,l}$, $\phi^c_{1,i,l}$ & \begin{tabular}[c]{@{}l@{}} $\theta^r_{1,i,l} = \theta^c_{1,i,l} + \pi$ \\  $\phi^r_{1,i,l} = \phi^c_{1,i,l} + \pi$ \end{tabular} \\ \hline
	Fading factor & $\widetilde{f}_{1,i,l}^c  \widetilde{F}_{1,i,l}^c e^{(-j\Delta \varphi_{1,i,l}^c)}$ & \begin{tabular}[c]{@{}l@{}} $\widetilde{f}_{1,i,l}^r  \widetilde{F}_{1,i,l}^r e^{(-j\Delta \varphi_{1,i,l}^r)}$ \\
		$\widetilde{f}_{2,i,0}^r  \widetilde{F}_{2,i,0}^r e^{(-j\Delta \varphi_{2,i,0}^r)}$ \\
	$\widetilde{f}_{3,i,0}^r  \widetilde{F}_{3,i,0}^r e^{(-j\Delta \varphi_{3,i,0}^r)}$
	\end{tabular}
	  \\ \hline
	Phase offset& $\Delta \varphi_{1,i,l}^c = \frac{2\pi f_c d_{c,1,i,l}}{c_0}$ &
	\begin{tabular}[c]{@{}l@{}} $\Delta \varphi_{1,i,l}^r = 2 \frac{2\pi f_c 2  d_{r,1,i,l}}{c_0} $ \\
		$\Delta \varphi_{2,i,l}^r = \frac{2\pi f_c  (d_{r,1,i,0} +d_{r,2,i,0})}{c_0} $ \\
		$\Delta \varphi_{3,i,l}^r = \frac{2\pi f_c  (d_{r,1,i,0} +d_{r,3,i,0})}{c_0}$
	 \end{tabular} \\ \hline
	\begin{tabular}[c]{@{}l@{}} {\color{black}Small-scale} \\ fading factor
	\end{tabular}	& $\widetilde{f}_{1,i,l}^c$ & $\widetilde{f}_{1,i,l}^r$, $\widetilde{f}_{2,i,0}^r$, $\widetilde{f}_{3,i,0}^r$ \\ \hline
	\begin{tabular}[c]{@{}l@{}} {\color{black}Large-scale}\\ fading factor
	\end{tabular}& $\widetilde{F}_{1,i,l}^c = \frac{\lambda^2}{(4\pi)^2 {d_i^2}}$ &
	\begin{tabular}[c]{@{}l@{}} $\widetilde{F}_{1,i,l}^r = \frac{\lambda^2}{(4\pi)^3 {d^4_{r,1,i,l}}}$ \\
		$\widetilde{F}_{2,i,l}^r = \frac{\lambda^2}{(4\pi)^3 {{d^2_{r,1,i,0}}{d^2_{r,2,i,0}}}}$ \\
		$\widetilde{F}_{3,i,l}^r = \frac{\lambda^2}{(4\pi)^3 {{d^2_{r,1,i,0}}{d^2_{r,3,i,0}}}}$
	\end{tabular}  \\ \hline
	\end{tabular}
\end{table}

\subsection{Received Signal Model of Users}\label{sec:system_model_1}
\subsubsection{Received Signal Model of Users}\label{sec:system_model_1_1}

The received signal of users can be expressed as
\begin{equation}\label{equ:UE_Signal_1}
{{{\bf Y}_{UE}}}  = {{\bf H}^T_1}{{\bf X}{_1}} + {{\bf H}^T_2}{{\bf X}{_2}} + {{\bf H}^T_3}{{\bf X}{_3}} + {{{\bf Z}_C}}, \in {{\mathcal C}^{3K \times L}},
\end{equation}
where ${{{\bf Z}_C}} \in {{\mathcal C}^{3K \times L}}$ is the Gaussian communication noise, the ISAC signal sent by ISAC $\rm{BS}_1$, ISAC $\rm{BS}_2$ and ISAC $\rm{BS}_3$ can be expressed as
\begin{equation}\label{equ:UE_Signal_2}
\begin{aligned}
{{\bf X}_{j=1,2,3}}  &={{\bf W}_j^T  {{\bf S}_j}}  \\
& =
\begin{cases}
	\sum_{i=1}^{K} &{\bf X}_{1,i} \in {\mathcal{C}}^{N_t \times L}  \\
	\sum_{i=1+K}^{2K} &{\bf X}_{2,i} \in {\mathcal{C}}^{N_t \times L}  \\
	\sum_{i=1+2K}^{3K} &{\bf X}_{3,i} \in {\mathcal{C}}^{N_t \times L}
\end{cases}
\end{aligned},
\end{equation}
where ${\bf W}_j \in {{\mathcal C}^{K\times N_t}}$ is the precoding of ISAC ${\rm{BS}}_j$, ${\bf S}_j \in {{\mathcal C}^{K\times L}} $ is the user data sent by ISAC ${\rm{BS}}_j$, ${\bf X}_{j,i} \in {\mathcal{C}}^{N_t \times L}$ is the $i$-th user signal sent by ISAC ${\rm{BS}}_j$,
$L$ is the length of data stream.

For each user, the desired signal refers to the data sent to the user by the connected ISAC base station, while interference signal includes multiuser interference signal and interference signal from the adjacent ISAC base station.
Then, the received signal of the $i$-th user can be expressed as
\begin{equation}
\begin{aligned}
& {{{\bf Y}_{UE_{i,1 \le i \le K}}}} =  {\bf h}^T_{1,i}{{\bf X}_{1,i}} + \\
& \qquad \quad \overbrace{{\bf h}^T_{2,i}{\bf X}_2 + {\bf h}^T_{3,i}{\bf X}_3 + {{\bf Z}_{C_i}} +\sum_{j=1,j \ne i}^{K} {\bf h}^T_{1,i}{{\bf X}_{1,j}}}^{\rm interference} \\
& {{{\bf Y}_{UE_{i,1+K \le i \le 2K}}}} =  {\bf h}^T_{2,i}{{\bf X}_{2,i}} + \\
& \qquad \quad \overbrace{{\bf h}^T_{1,i}{\bf X}_1 + {\bf h}^T_{3,i}{\bf X}_3 + {{\bf Z}_{C_i}} +\sum_{j=1+K,j \ne i}^{2K} {\bf h}^T_{2,i}{{\bf X}_{2,j}}}^{\rm interference} \\
& {{{\bf Y}_{UE_{i,1+2K \le i \le 3K}}}} = {\bf h}^T_{3,i}{{\bf X}_{3,i}} + \\
& \qquad \quad \overbrace{{\bf h}^T_{1,i}{\bf X}_1 +  {\bf h}^T_{2,i}{\bf X}_2 + {{\bf Z}_{C_i}} +\sum_{j=1+2K,j \ne i}^{3K} {\bf h}^T_{3,i}{{\bf X}_{3,j}}}^{\rm interference}
\end{aligned}.
\label{equ:UE_Signal_3}
\end{equation}

\subsubsection{Received Signal Model of ISAC Base Stations}\label{sec:system_model_1_2}

For each ISAC base station, the desired echo signal refers to its own echo signal reflected by targets, the interference echo signal includes the echo signal reflected by scatterers and the echo signal of adjacent ISAC base stations reflected by targets.
Then, the received signal of ISAC ${\rm BS}_1$ can be expressed as
\begin{equation}
\begin{aligned}
{{{\bf Y}_{BS_1}}} = & {{\bf G}^T_{1,1,0}}{{\bf X}{_1}} + \\
& \overbrace{ {{\bf G}^T_{1,2}}{{\bf X}{_2}} + {{\bf G}^T_{1,3}}{{\bf X}{_3}} + {{\bf Z}_{R_1}} + \sum_{l=1}^{L_{p}-1} {{{\bf G}^T_{1,1,l}}{{\bf X}{_1}}}}^{\rm interference}
\end{aligned},
\label{equ:BS_Signal_1}
\end{equation}
%
where ${{{\bf Z}_{R_1}}} \in {{\mathcal C}^{3K \times L}}$ is the Gaussian radar noise and ground clutter \cite{[clutter]}, ${{\bf G}_{1,1,0}}$ is the echo path from targets and ${{\bf G}_{1,1,l;l\ne 0}}$ is the $l$-th echo path from scatterers, which can be expressed as
\begin{equation}\label{equ:BS_Signal_2}
\begin{aligned}
{{\bf G}_{1,1,l}} = [{\bf g}_{1,1,1,l},{\bf g}_{1,1,2,l},…,{\bf g}_{1,1,3K,l}], \in {{\mathcal C}^{N_t \times 3K }}, 0 \le l \le L_p \\
\end{aligned}.
\end{equation}
The received signal of ISAC ${\rm BS}_2$ and ISAC ${\rm BS}_3$ can be expressed in a similar way.
%

\section{Optimization Problem Formulation for Collaborative Precoding Design}\label{sec:OPF}
The aim of collaborative precoding design is to maximizing the SINR of the received signals of the users and ISAC base stations subject to some additional constraints.
In general, the optimization problem formulation for collaborative precoding design can be expressed as
\begin{equation}
	\begin{aligned}
		{\mathcal{P}} : \min_{{{\bf T}_j}} & \quad P_t  \\
		s.t.
		& \quad {\gamma_{R_j}}   \ge {\zeta }_{R_j}, j=1,2,3
		\\
		& {\gamma_{C_i}}   \ge {\zeta }_{C_i} \quad  or \quad {\gamma_{C_i}^{CI}}   \ge {\zeta }_{C_i}, 1 \le i \le 3K \\
		& \quad {{\bf T}_{1,i}} \succeq 0, {{\bf T}_{1,i}} = {{\bf T}_{1,i}^H} \\
		& \quad  rank({{\bf T}_{1,i}}) = 1, \forall i \\
		& {\rm CI \quad constraint } \quad \eqref{equ:OWD_CI_2}\\
		& {\rm TPC \quad constraint } \quad \eqref{equ:OWDC_6} \quad or\\
		& {\rm PPC \quad constraint } \quad \eqref{equ:OWDC_7} \quad or\\
		& {\rm CMC \quad and \quad SC \quad constraint } \quad \eqref{equ:OWDC_8}
	\end{aligned},
	\label{equ:JOA}
\end{equation}
{\color{black}
where ${{\bf T}_{1,i}} = {{\bf x}_{1,i}} {{\bf x}_{1,i}^H} \succeq 0, \in {{\mathcal C}^{N_t \times N_t}} $ with ${{\bf x}_{1,i}} \in {{\mathcal C}^{N_t \times 1}}$ being the received signal of the $i$-th user.
Since ${{\bf x}_{1,i}} \ne \bf 0$ is a vector, $rank({{\bf x}_{1,i}}) = 1$.
Then, $rank({{\bf T}_{1,i}}) = 1$ can be deduced by the rank theorem for the product of matrices:

\textit{if $\bf A$ = $\bf B C $, then $rank( {\bf A} ) \le \min (rank( {\bf B} ), rank( {\bf C} )) $}.

}

Details of the collaborative precoding design criterions will be introduced in section \ref{sec:OWDC}.

\subsection{Collaborative Precoding Design Criterion}\label{sec:OWDC}
\subsubsection{SINR Constraint}\label{sec:OPF_1}

The detection probability of a target is usually a monotonically increasing function of SINR of ISAC base stations. The communication performance of UEs is also a monotonically increasing function of SINR of UEs.
In order to meet specific communication and sensing requirements, communication and sensing SINR should meet certain requirements, referred to as SINR constraint.

Based on the received signal model of users and ISAC base stations, introduced in section \ref{sec:system_model_1}, the SINR of the $i$-th user can be expressed as
\begin{equation}\label{equ:OWDC_1}
\begin{aligned}
& {\gamma_{C_{i,1 \le i \le K}}} \\
& = \frac{ |{\bf h}^T_{1,i}{{\bf X}_{1,i}}|^2 } { |{\bf h}^T_{2,i}{\bf X}_2|^2 + |{\bf h}^T_{3,i}{\bf X}_3|^2 + {{\bf \sigma}^2_{C_i}} +\sum_{j=1,j \ne i}^{K} |{\bf h}^T_{1,i}{{\bf X}_{1,j}}|^2 } \\
& = \frac{ {\rm tr}({\bf h}_{1,i}^{*} {\bf h}_{1,i}^T {{\bf T}_{1,i}})  }   { {\rm tr}({\bf h}_{2,i}^{*} {\bf h}_{2,i}^T {{\bf T}_{2}} ) + {\rm tr}({\bf h}_{3,i}^{*} {\bf h}_{3,i}^T {{\bf T}_{3}} )} \\
& + \frac{ {\rm tr}({\bf h}_{1,i}^{*} {\bf h}_{1,i}^T {{\bf T}_{1,i}})  } { \sum_{j=1,j \ne i}^{K} {\rm tr}({\bf h}_{1,i}^{*} {\bf h}_{1,i}^T {{\bf T}_{1,j}})    +  {{\bf \sigma}^2_{C_i}} }
\end{aligned},
\end{equation}
where
\begin{equation}\label{equ:OWDC_2}
\begin{aligned}
{{\bf T}_{j=2,3}} &= {\bf X}_j {\bf X}_j^H \succeq 0, \in {{\mathcal C}^{N_t \times N_t} }\\
rank ({{\bf T}_{1,i}}) &= 1
\end{aligned},
\end{equation}
are the autocorrelation matrix of ISAC transmitting signal,
${\rm tr}(\bf X)$ denotes the trace of $\bf X$,  ${\bf \sigma}^2_{C_i}$ is the power of the communication noise of the $i$-th user.
The SINR of the other users, i.e. $ K+1 \le i \le 3K$, can be expressed in a similar way.
Similarly, the SINR of ISAC ${\rm BS}_1$ can be expressed as
\begin{equation}\label{equ:OWDC_3}
\begin{aligned}
{\gamma_{R_1}} &= \frac{ |{{\bf G}^T_{1,1,0}}{{\bf X}_1}|^2  }{ |{{\bf G}^T_{1,2}}{{\bf X}{_2}}|^2 + |{{\bf G}^T_{1,3}}{{\bf X}{_3}}|^2 + {{\bf \sigma}_{R_1}^2} + \sum_{l=1}^{L_{p}-1} |{{{\bf G}^T_{1,1,l}}{{\bf X}{_1}}}|^2 } \\
&= \frac { {\rm tr}({{\bf G}{_{1,1,0}^{*}}}{{\bf G}^T_{1,1,0}} {{\bf T}_{1}} )}   { {\rm tr}({{\bf G}_{1,2}^{*}}{{\bf G}^T_{1,2}} {{\bf T}_{2}} ) + {\rm tr}({{\bf G}_{1,3}^{*}}{{\bf G}^T_{1,3}} {{\bf T}_{3}} ) } \\
&+ \frac { {\rm tr}({{\bf G}{_{1,1,0}^{*}}}{{\bf G}^T_{1,1,0}} {{\bf T}_{1}} )}   { \sum_{l=1}^{L_{p}-1}  {\rm tr}({{\bf G}{_{1,1,l}^{*}}}  {{\bf G}{_{1,1,l}^T}} {{\bf T}_{1}} ) + {\bf \sigma}_{R_1}^2  }
\end{aligned},
\end{equation}
where ${{\bf \sigma}_{R_1}^2}$ is the power of the radar channel noise, the autocorrelation matrix of ${\bf X}{_1}$ is
\begin{equation}\label{equ:OWDC_4}
{{\bf T}_{1}} = {\bf X}{_1} {\bf X}{_1}^H \succeq 0, \in {{\mathcal C}^{N_t \times N_t}}.
\end{equation}
The SINR of ISAC ${\rm BS}_2$ and ISAC ${\rm BS}_3$, i.e. $ {\gamma_{R_2}}$ and ${\gamma_{R_3}}$, can be expressed in a similar way.
Then the SINR constraint can be expressed as
\begin{equation}
\begin{aligned}
  {\gamma_{R_j}}   &\ge {\zeta }_{R_j}, j=1,2,3 \\
  {\gamma_{C_i}}   &\ge {\zeta }_{C_i}, 1 \le i \le 3K
\end{aligned},
\label{equ:OWDC_5}
\end{equation}
where ${\zeta }_{R_j}$ is the SINR threshold of ISAC ${\rm BS}_j$, ${\zeta }_{C_i}$ is the SINR threshold of the $i$-th user.

%

\subsubsection{Transmit Power Constraint}\label{sec:OPF_2}


In order to reduce spectrum interference in wireless environment, communication and radar sensing have transmit power constraint, including total transmit power constraint (TPC) and per-antenna transmit power constraint (PPC).
%
TPC can be expressed as
\begin{equation}
\begin{aligned}
& \quad {\frac{1}{L}  {|| {{\bf X}_j} ||}_F^2 } \le P_t, j = 1,2,3
\end{aligned},
\label{equ:OWDC_6}
\end{equation}
where $P_t$ is the maximum value of transmit power.
%
PPC can be expressed as
\begin{equation}
\begin{aligned}
& \quad \frac{1}{L}  {diag({{\bf X}_j}{{\bf X}_j}^H)} \le \frac{P_t}{N_t} {\bf I}_{N_t}, j = 1,2,3
\end{aligned},
\label{equ:OWDC_7}
\end{equation}
where ${\bf I}_{N_t}$ denotes the $ {N_t} \times {N_t} $ identity matrix.

\subsubsection{CMC and SC}\label{sec:OPF_4}

CMC is to enforce the modulus of each element $x(k)$ of the signal ${\bf X}$ to be a constant.
In order to achieve a trade-off between the output SINR and other desired waveform characteristics (i.e., pulse compression and ambiguity) \cite{[ambiguity_21]}, radar usually requires the waveform to satisfy the constant modulus and similarity constraints, optimizing the sensing performance in the appropriate neighborhood of reference waveforms known to have good characteristics.
To satisfy the worst-case similarity optimum, CMC and SC can be expressed as \cite{[SC_37]}
\begin{equation}
\begin{aligned}
||{{\bf X}_j} - {{\bf X}_0}||_\infty \le {\epsilon }_{j}, j= 1,2,3
\end{aligned},
\label{equ:OWDC_8}
\end{equation}
where $||\bf X||_\infty$ means the max of $\bf X$, ${\epsilon }_{j}$ is the threshold of the difference between ${{\bf X}_j}$ and ${\bf X}_0$, ${{\bf X}_0} \in \mathcal{C}^{N_t * L}$ is a known benchmark radar signal matrix that has constant modulus entries, e.g., orthogonal
chirp signals \cite{[Liu_1],[Liu_38]}.

%
%
%
%


\subsection{Collaborative Precoding Design}\label{sec:OWD_DI}

It should be noted that the traditional precoding is designed for ${\bf W}$. In this paper, an overall design is considered for signal ${\bf X}$ after precoding, which can be regarded as a generalized precoding design \cite{[Liu_1]}.
\begin{table}[h]
	\caption{Optimization Problems for Collaborative Precoding Design}
    \centering
	\label{label:OP_DI}
	\begin{tabular}{l|l}
		\hline \hline
		 \begin{tabular}[c]{@{}l@{}} Optimization problem \\
			with DI and TPC \end{tabular}   & \begin{tabular}[c]{@{}l@{}}
			{\color{black} ${\mathcal{P}^{\mathsf {DI}}_{\mathsf {TPC}}}$} : $\min_{{{\bf X}_j}}$ $P_t$
			\\
			s.t. \quad ${\gamma_{R_j}}   \ge {\zeta }_{R_j}, j=1,2,3$
			\\ \quad \quad${\gamma_{C_i}}   \ge {\zeta }_{C_i}, 1 \le i \le 3K$
			\\ \quad \quad $\frac{1}{L} ||{{\bf X}_j}||_F^2  \le {P_t},j=1,2,3$  \end{tabular} \\ \hline
		\begin{tabular}[c]{@{}l@{}} Optimization problem \\
			with DI and PPC \end{tabular} & \begin{tabular}[c]{@{}l@{}}
			{\color{black} ${\mathcal{P}^{\mathsf {DI}}_{\mathsf {PPC}}}$ } : $\min_{{{\bf X}_j}}$ $P_t$ \\
			s.t.
			\quad ${\gamma_{R_j}}   \ge {\zeta }_{R_j}, j=1,2,3$
			\\ \quad \quad${\gamma_{C_i}}   \ge {\zeta }_{C_i}, 1 \le i \le 3K$ \\
			\quad \quad $\frac{1}{L}  {diag({{\bf X}_j}{{\bf X}_j}^H)} \le \frac{P_t}{N_t} {\bf I}_{N_t},j=1,2,3$
		\end{tabular} \\ \hline
		\begin{tabular}[c]{@{}l@{}} Optimization problem \\
			with DI and CMC \end{tabular} & \begin{tabular}[c]{@{}l@{}}
			{\color{black} ${\mathcal{P}^{\mathsf {DI}}_{\mathsf {CMC}}}$} : $\min_{{{\bf X}_j}}$ $P_t$ \\
			s.t.
			\quad ${\gamma_{R_j}}   \ge {\zeta }_{R_j}, j=1,2,3$
			\\ \quad \quad${\gamma_{C_i}}   \ge {\zeta }_{C_i}, 1 \le i \le 3K$ \\
			\quad \quad $||{{\bf X}_j} - {{\bf X}_0}||_\infty \le {\epsilon }_{j}, j= 1,2,3$
		\end{tabular} \\ \hline
	\end{tabular}
\end{table}
Then, the optimization problems for collaborative precoding design can be formulated in Table \ref{label:OP_DI}.

\subsection{Collaborative Precoding Design with CI}\label{sec:OWD_CI}

\begin{figure}[ht]
	\includegraphics[scale=0.25]{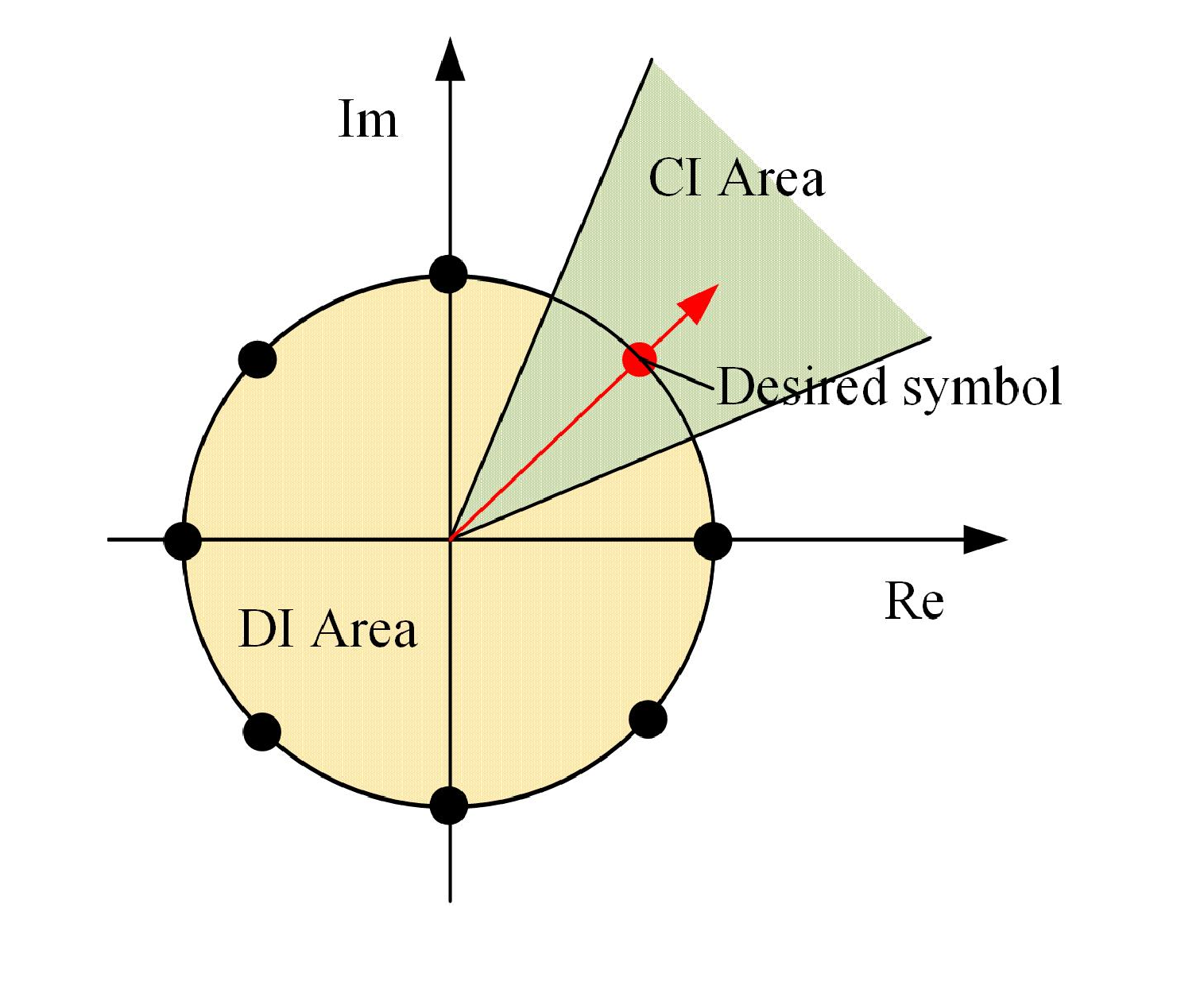}
	\centering
	\caption{CI and DI area for desired symbols.}
	\label{fig:CI}
\end{figure}

In section \ref{sec:OWD_DI}, we treat multiuser interference as DI that needs to be suppressed and therefore put it in the denominator of SINR. It should be noted that the multiuser interference is also known to the ISAC base station because the base station knows the communication channel matrix and the precoding matrix. In this section, we use the CI for collaborative precoding to achieve the purpose of co-existence of adjacent ISAC base stations.
As shown in Fig. \ref{fig:CI}, the yellow area is the DI area, the green area is the CI area for the desired symbol, which means that the received signal in the CI area can be demodulated correctly.
For simplicity of description, this paper derives the CI conversion problem for $K$ users served by ISAC ${\rm BS}_1$, and similar derivations can be used for other users.
Assuming that all multiuser interference is converted to CI, SINR of users can be greatly improved, which can be expressed as
\begin{equation}\label{equ:OWD_CI_1}
\begin{aligned}
{\gamma^{CI}_{C_i}} = & \frac{ {\rm tr}({\bf h}_{1,i}^{*} {\bf h}^T_{1,i} {{\bf T}_{1}})  }   { {\rm tr}({\bf h}_{2,i}^{*} {\bf h}^T_{2,i} {{\bf T}_{2}} ) + {\rm tr}({\bf h}_{3,i}^{*} {\bf h}^T_{3,i} {{\bf T}_{3}} )   +  {{\bf \sigma}^2_{C_i}} }, 1 \le i \le K
\end{aligned}.
\end{equation}
According to \cite{[CI_29], [CI_35]}, the proposed concept of interference utilization has been shown to be beneficial to a variety of modulation formats, such as phase shift keying (PSK).
%
The PSK symbol can be denoted as $e^{j\phi_k}$. Then the CI constraint can be expressed as \cite{[Liu_1]}
\begin{equation}
\begin{aligned}
 |{\rm Im}& ({\bf h}_{1,i}^T  \sum_{k=1}^{K} {\bf T}_{1,k} e^{j(\phi_k - \phi_i)})|^2 \le \\
& ({\rm Re}({\bf h}_{1,i}^T \sum_{k=1}^{K} {\bf T}_{1,k} e^{j(\phi_k - \phi_i)}) - \widetilde{\zeta}_i){\tan}\psi\\
 \widetilde{\zeta}_i & = {\zeta}_{C_i}({\bf h}_{1,i}^{*} {\bf h}_{1,i}^T  {\bf T}_2  + {{\bf \sigma}^2_{C_i}}), , 1 \le i \le K
\end{aligned},
\label{equ:OWD_CI_2}
\end{equation}
where $\psi = \frac{\pi}{M_p}$, $M_p$ is the PSK modulation order, ${\zeta}_i$ is the SINR threshold of the $i$-th user, ${\rm Im}({\bf x})$ and ${\rm Re}({\bf x})$ denote the imaginary and real parts of $\bf x$, respectively.
Then, considering TPC, PPC and CMC respectively, optimization problems for collaborative precoding design with CI can be formulated in Table \ref{label:OP_CI}.

\begin{table}[ht]
	\caption{Optimization Problems for Collaborative Precoding Design with CI}
    \centering
	\label{label:OP_CI}
	\begin{tabular}{l|l}
		\hline \hline
		\begin{tabular}[c]{@{}l@{}} Optimization problem \\
			with CI and TPC \end{tabular} & \begin{tabular}[c]{@{}l@{}}
			{\color{black} ${\mathcal{P}^{\mathsf {CI}}_{\mathsf {TPC}}}$} : $\min_{{{\bf X}_j}}$ $P_t$
			\\
			s.t. \quad ${\gamma_{R_j}}   \ge {\zeta }_{R_j}, j=1,2,3$
			\\ \quad \quad${\gamma^{CI}_{C_i}}   \ge {\zeta }_{C_i}, 1 \le i \le 3K$
			\\ \quad \quad $\frac{1}{L} ||{{\bf X}_j}||_F^2  \le {P_t},j=1,2,3$  \\
			$|{\rm Im}({\bf h}_{1,i}^T \sum_{k=1}^{K} {\bf T}_{1,k} e^{j(\phi_k - \phi_i)})|^2 \le$ \\
			$ ({\rm Re}({\bf h}_{1,i}^T \sum_{k=1}^{K} {\bf T}_{1,k} e^{j(\phi_k - \phi_i)}) - \widetilde{\zeta}_i){\tan}\psi$
		\end{tabular}\\ \hline
		\begin{tabular}[c]{@{}l@{}} Optimization problem \\
			with CI and PPC \end{tabular} & \begin{tabular}[c]{@{}l@{}}
			{\color{black} ${\mathcal{P}^{\mathsf {CI}}_{\mathsf {PPC}}}$} : $\min_{{{\bf X}_j}}$ $P_t$
			\\
			s.t. \quad ${\gamma_{R_j}}   \ge {\zeta }_{R_j}, j=1,2,3$
			\\ \quad \quad${\gamma^{CI}_{C_i}}   \ge {\zeta }_{C_i}, 1 \le i \le 3K$
			\\ \quad \quad $\frac{1}{L} ||{{\bf X}_j}||_F^2  \le {P_t},j=1,2,3$  \\
			$|{\rm Im}({\bf h}_{1,i}^T \sum_{k=1}^{K} {\bf T}_{1,k} e^{j(\phi_k - \phi_i)})|^2 \le$ \\
			$ ({\rm Re}({\bf h}_{1,i}^T \sum_{k=1}^{K} {\bf T}_{1,k} e^{j(\phi_k - \phi_i)}) - \widetilde{\zeta}_i){\tan}\psi$
		\end{tabular}\\ \hline
		\begin{tabular}[c]{@{}l@{}} Optimization problem \\
			with CI and CMC \end{tabular} & \begin{tabular}[c]{@{}l@{}}
			{\color{black} ${\mathcal{P}^{\mathsf {CI}}_{\mathsf {CMC}}}$} : $\min_{{{\bf X}_j}}$ $P_t$ \\
			s.t.
			\quad ${\gamma_{R_j}}   \ge {\zeta }_{R_j}, j=1,2,3$
			\\ \quad \quad${\gamma^{CI}_{C_i}}   \ge {\zeta }_{C_i}, 1 \le i \le 3K$ \\
			\quad \quad $||{{\bf X}_j} - {{\bf X}_0}||_\infty \le {\epsilon }_{j}, j= 1,2,3$ \\
			$|{\rm Im}({\bf h}_{1,i}^T \sum_{k=1}^{K} {\bf T}_{1,k} e^{j(\phi_k - \phi_i)})|^2 \le$ \\
			$ ({\rm Re}({\bf h}_{1,i}^T \sum_{k=1}^{K} {\bf T}_{1,k} e^{j(\phi_k - \phi_i)}) - \widetilde{\zeta}_i){\tan}\psi$
		\end{tabular}\\ \hline
	\end{tabular}
\end{table}

\section{Collaborative Precoding Design}\label{sec:OWD}

This section introduces two collaborative precoding design algorithms, JOA and SOA.
Since, the solutions of {\color{black} ${\mathcal{P}^{\mathsf {DI}}_{\mathsf {TPC}}}$}, {\color{black} ${\mathcal{P}^{\mathsf {DI}}_{\mathsf {PPC}}}$}, {\color{black} ${\mathcal{P}^{\mathsf {DI}}_{\mathsf {CMC}}}$}, {\color{black} ${\mathcal{P}^{\mathsf {CI}}_{\mathsf {TPC}}}$}, {\color{black} ${\mathcal{P}^{\mathsf {CI}}_{\mathsf {PPC}}}$}, {\color{black} ${\mathcal{P}^{\mathsf {CI}}_{\mathsf {CMC}}}$} are similar, we solve the optimization problem {\color{black} ${\mathcal{P}^{\mathsf {DI}}_{\mathsf {PPC}}}$} as an example. The differences in the solution of each problem are listed in Table \ref{alg:six_JOA}.

\subsection{JOA for Collaborative Precoding Design}\label{sec:JOA}

Based on \eqref{equ:OWDC_1} and \eqref{equ:OWDC_3}, {\color{black} ${\mathcal{P}^{\mathsf {DI}}_{\mathsf {PPC}}}$} can be rewritten as
\begin{equation}
\begin{aligned}
{\color{black} {\mathcal{P}^{\mathsf {DI}}_{\mathsf {PPC}}}} : \min_{{{\bf T}_j}} & \quad P_t  \\
s.t.
& \quad {\gamma_{R_j}}   \ge {\zeta }_{R_j}, j=1,2,3
\\
& \quad \quad {\gamma_{C_i}}   \ge {\zeta }_{C_i}, 1 \le i \le 3K \\
& \quad {{\bf T}_{1,i}} \succeq 0, {{\bf T}_{1,i}} = {{\bf T}_{1,i}^H} \\
& \quad  rank({{\bf T}_{1,i}}) = 1, \forall i \\
& \quad \frac{1}{L}  {diag({{\bf T}_j})} \le \frac{P_t}{N_t} {\bf I}_{N_t}, j = 1,2,3 \\
\end{aligned}.
\label{equ:JOA_1}
\end{equation}
Since \eqref{equ:OWDC_1} and \eqref{equ:OWDC_3} are not standard convex equation, we need to rewrite the SINR constraint.
The SINR constraint transformation is similar in different base station systems. For simple description, ISAC ${\rm BS}_1$ is taken as an example, i.e., $j = 1, 1 \le i \le K$.
Then, the SINR constraint of ISAC ${\rm BS}_1$ can be rewritten as
\begin{equation}\label{equ:JOA_2}
\begin{aligned}
{\rm tr}({\bf h}_{1,i}^{*} {\bf h}^T_{1,i} {{\bf T}_{1,i}}) &\ge {\zeta }_{C_i}( {\rm tr}({\bf h}_{2,i}^{*} {\bf h}^T_{2,i} {{\bf T}_{2}} ) + {\rm tr}({\bf h}_{3,i}^{*} {\bf h}^T_{3,i} {{\bf T}_{3}} ) + \\
&\sum_{j=1,j \ne i}^{K} {\rm tr}({\bf h}_{1,i}^{*} {\bf h}^T_{1,i} {{\bf T}_{1,j}})    +  {{\bf \sigma}^2_{C_i}}   )
\end{aligned},
\end{equation}
\begin{equation}\label{equ:JOA_3}
\begin{aligned}
 {\rm tr}({{\bf G}_{1,1,0}^{*}}{{\bf G}{_{1,1,0}^T}} {{\bf T}_{1}} ) &\ge  {\zeta }_{R_1} ({\rm tr}({{\bf G}_{1,2}^{*}}{{\bf G}{_{1,2}^T}} {{\bf T}_{2}} ) +  {\rm tr}({{\bf G}_{1,3}^{*}}{{\bf G}{_{1,3}^T}} {{\bf T}_{3}} ) \\ +  &\sum_{l=1}^{L_{p}-1}  {\rm tr}({{\bf G}_{1,1,l}^{*}}  {{\bf G}{_{1,1,l}^T}} {{\bf T}_{1}} ) + {\bf \sigma}_{R_1}^2)
\end{aligned}.
\end{equation}
Due to the constraint of $rank({{\bf T}_{1,i}}) = 1$, the problem {\color{black} ${\mathcal{P}^{\mathsf {DI}}_{\mathsf {PPC}}}$} is still non-convex.
We can transform {\color{black}${\mathcal{P}^{\mathsf {DI}}_{\mathsf {PPC}}}$} into a standard semidefinite program (SDP) problem by omitting the constraint of $rank({{\bf T}_{1,i}}) = 1$.
Then, a suboptimal solution ${\bf T}_j$ can be obtained by solving the SDP problem \cite{[SDP_37],[SDP_92]}.
Furthermore, an approximate solution can be obtained by using standard rank-1 approximation techniques such as eigenvalue decomposition \cite{[liu_seperation_38]}.
The diagonal matrix ${\bf U}_j \in {{\mathcal C}^{N_t \times N_t}} $ and the square matrix of eigenvectors ${\bf V}_j \in {{\mathcal C}^{N_t \times N_t}} $ can be obtained by the eigenvalue decomposition of ${\bf T}_j$.
The elements of ${\bf U}_j$ are the eigenvalues, the eigenvectors of ${\bf V}_j$ are sorted according to the eigenvalues from large to small.
Extract the first $L$ eigenvectors to get the matrix ISAC signal matrix ${\bf X}_j \in {{\mathcal C}^{N_t \times L}}$.

The proposed optimization algorithm JOA can be summarized by the following Algorithm \ref{alg:JOA_1}. JOA for solving the other 5 problems has been listed in Table \ref{alg:six_JOA}.
\begin{algorithm}[htb]
	\caption{Joint optimization algorithm (JOA)}
	\label{alg:JOA_1}
	\begin{algorithmic}
		\Require
		$\bf H$, ${\zeta }_{C,i}$, ${\zeta }_{R_j}$ and $N_t$;
		\State \quad 1). According to the relationship between communication and radar sensing channel parameters listed in table \ref{label:ISAC_CSI}, generate $\bf G$ by \eqref{equ:Rad_CSI_1} and \eqref{equ:Rad_CSI_2};
		\State \quad 2). Rewrite the SINR constraint by \eqref{equ:JOA_2} and \eqref{equ:JOA_3};
		\State \quad 3). Transform {\color{black} ${\mathcal{P}^{\mathsf {DI}}_{\mathsf {PPC}}}$} into the SDP problem by omitting the rank-1 constraint;
		\State \quad 4). Obtain ${\bf T}_j$ by solving the SDP problem;
		\State \quad 5). Obtain ${\bf U}_j $ and $ {\bf V}_j $ by eigenvalue decomposition of ${\bf T}_j$;
		\State \quad 6). Extract the first $L$ eigenvectors to get ${\bf X}_j$
		\Ensure ${\bf X}_j, j=1,2,3$.
	\end{algorithmic}
\end{algorithm}

\begin{table}[h]
	\caption{JOA and SOA for solving six optimization problems.}
    \centering
	\label{alg:six_JOA}
	\begin{tabular}{l|l|l}
		\hline
		\hline
		Problems & Constraints & Main steps \\ \hline
		{\color{black} ${\mathcal{P}^{\mathsf {DI}}_{\mathsf {TPC}}}$} & DI and TPC &  \begin{tabular}[c]{@{}l@{}} $\cdot$ Relax the problem into the SDP problem \\ \quad by omitting the rank-1 constraint. \end{tabular}    \\ \hline
		{\color{black} ${\mathcal{P}^{\mathsf {DI}}_{\mathsf {PPC}}}$} & DI and PPC & \begin{tabular}[c]{@{}l@{}} $\cdot$ Divide the diagonal constraint of TPC into \\ \quad N quadratic equality constraints. \\
			$\cdot$ Relax the problem into the SDP problem \\ \quad by omitting the rank-1 constraint. \end{tabular} \\ \hline
		{\color{black} ${\mathcal{P}^{\mathsf {DI}}_{\mathsf {CMC}}}$} & DI and CMC & \begin{tabular}[c]{@{}l@{}} $\cdot$ Solve the NP-hard problem based on a \\ \quad general framework of the BnB  \\\quad methodology \cite{[BnB_57],[Liu_1]}. \end{tabular} \\ \hline
		{\color{black} ${\mathcal{P}^{\mathsf {CI}}_{\mathsf {TPC}}}$} & CI and TPC & \begin{tabular}[c]{@{}l@{}} $\cdot$ Derive the real representation of \\ \quad the problem {\color{black} ${\mathcal{P}^{\mathsf {CI}}_{\mathsf {TPC}}}$} \cite{[Liu_parameters_1]}. \\
			$\cdot$ Relax the problem into the SDP problem \\ \quad by omitting the rank-1 constraint. \end{tabular}   \\ \hline
		{\color{black} ${\mathcal{P}^{\mathsf {CI}}_{\mathsf {PPC}}}$} & CI and PPC & \begin{tabular}[c]{@{}l@{}} $\cdot$ Derive the real representation of \\ \quad the problem {\color{black} ${\mathcal{P}^{\mathsf {CI}}_{\mathsf {PPC}}}$}. \\
			$\cdot$ Divide the diagonal constraint of TPC into \\ \quad N quadratic equality constraints. \\
			$\cdot$ Relax the problem into the SDP problem \\ \quad by omitting the rank-1 constraint. \end{tabular} \\ \hline
		{\color{black} ${\mathcal{P}^{\mathsf {CI}}_{\mathsf {CMC}}}$} & CI and CMC & \begin{tabular}[c]{@{}l@{}} $\cdot$ Derive the real representation of \\ \quad the problem {\color{black} ${\mathcal{P}^{\mathsf {CI}}_{\mathsf {CMC}}}$}. \\
			 $\cdot$ Solve the NP-hard problem based on \\ \quad a general framework of \\ \quad the BnB methodology. \end{tabular} \\ \hline
	\end{tabular}
\end{table}

\subsection{SOA for Collaborative Precoding Design}\label{sec:SOA}

Although JOA can transform the optimization problem {\color{black} ${\mathcal{P}^{\mathsf {DI}}_{\mathsf {PPC}}}$} into a convex problem and solve it by MATLAB's cvx tool, the complexity of this algorithm is high.
{\color{black}
Based on \eqref{equ:JOA_1}, the optimization problem ${\mathcal{P}^{\mathsf {DI}}_{\mathsf {PPC}}}$ contains multiple optimization objectives: maximizing the SINR of the radar sensing echo signal at the base station and maximizing the SINR of the received signal for user communication.
Maximizing the SINR of the base station radar echo signal can in turn be achieved by two subtasks: minimizing multipath interference signals (MPI) and minimizing echo interference from other ISAC base stations (MBI\_R).
Similarly, maximizing the SINR of the received signal for user communication can also be achieved by two subtasks: minimizing multiuser interference (MUI) and minimizing and interference from other ISAC base stations (MBI\_C).
Therefore, the optimization problem ${\mathcal{P}^{\mathsf {DI}}_{\mathsf {PPC}}}$ can be regarded as a multi-objective optimization problem \cite{[CI]}.

The iteration algorithm is a typical algorithm to solve the multi-objective optimization problem \cite{[ADMM]}.
Therefore, we propose SOA based on the iteration algorithm, that decompose the optimization problem into four subproblems, get the local optimal solution by solving the subproblems, and finally obtain the global optimal solution iteratively by the gradient descent method \cite{[CGM]}. Details of the iteration in SOA are introduced in Algorithm \ref{alg:SOA_1}.
}

SOA for different ISAC base stations is similar. For simple description, ISAC ${\rm BS}_1$ is taken as an example, i.e. $j = 1, 1 \le i \le K$.
According to section \ref{sec:OPF}, we can know that the interference to users contains two parts: MUI and MBI\_C.
The interference to ISAC base stations also consists of two parts: MPI and MBI\_R.
In order to decompose the optimization problem into four subproblems, we can rewrite the reciprocal of SINR of the $i$-th user as follow
\begin{equation}
\begin{aligned}
\frac{1}{\gamma_{C_i}} &=
\overbrace{ \frac{ {\rm tr}({\bf h}_{2,i}^{*} {\bf h}^T_{2,i} {{\bf T}_{2}} ) + {\rm tr}({\bf h}_{3,i}^{*} {\bf h}^T_{3,i} {{\bf T}_{3}} )} {{ {\rm tr}({\bf h}_{1,i}^{*} {\bf h}^T_{1,i} {{\bf T}_{1,i}})  }}}^{\rm MBI \_C}  \\ &+ \overbrace{\frac{	\sum_{j=1,j \ne i}^{K} {\rm tr}({\bf h}_{1,i}^{*} {\bf h}^T_{1,i} {{\bf T}_{1,j}})    +  {{\bf \sigma}^2_{C_i}}}
  { {\rm tr}({\bf h}_{1,i}^{*} {\bf h}^T_{1,i} {{\bf T}_{1,i}})  }}^{\rm noise + MUI}
\end{aligned},
\label{equ:SOA_1}
\end{equation}
where the first part indicates MBI\_C, and the second part indicates the noise and MUI.
Similarly, the reciprocal of SINR of ISAC $\rm BS_1$ can be rewritten as
\begin{equation}
\begin{aligned}
\frac{1}{\gamma_{R_1}} = &
\overbrace{\frac  { {\rm tr}({{\bf G}_{1,2}^{*}}{{\bf G}^T_{1,2}} {{\bf T}_{2}} ) + {\rm tr}({{\bf G}_{1,3}^{*}}{{\bf G}^T_{1,3}} {{\bf T}_{3}} )  } { {\rm tr}({{\bf G}_{1,1,0}^{*}}{{\bf G}{_{1,1,0}^T}} {{\bf T}_{1}} )}}^{\rm MBI \_R}  \\
+ & \overbrace{\frac  {  \sum_{l=1}^{L_{p}-1}  {\rm tr}({{\bf G}_{1,1,l}^{*}}  {{\bf G}{_{1,1,l}^T}} {{\bf T}_{1}} ) + {\bf \sigma}_{R_1}^2  } { {\rm tr}({{\bf G}_{1,1,0}^{*}}{{\bf G}{_{1,1,0}^T}} {{\bf T}_{1}} )}}^{\rm noise + MPI}
\end{aligned},
\label{equ:SOA_2}
\end{equation}
where the first part indicates MBI\_R, and the second part indicates the noise and MPI.
Then, we can decompose the problem ${\mathcal{P}_{2}}$ into four subproblems as follows.
\subsubsection{MUI to users}\label{sec:SOA_1}

Considering MUI and noise, the optimization problem is still non-convex because of the constraint of $rank({{\bf T}_{1,i}}) = 1$. By omitting the constraint of $rank({{\bf T}_{1,i}}) = 1$, the optimization problem can be transformed into a standard SDP problem
\begin{equation}
\begin{aligned}
{\mathcal{P}_{\mathsf {MUI}}} : \min_{{{\bf X}_1}} & \quad  \frac{N_t}{L} {diag({{\bf X}_1}{{\bf X}_1}^H)} \\
s.t.
& \frac{{{\bf \sigma}^2_{C_i}} + \sum_{j=1,j \ne i}^{K} {\rm tr}({\bf A}_{1,i} {{\bf T}_{1,j}} )}{ {\rm tr}({\bf A}_{1,i} {{\bf T}_{1,i}}) } \le \frac{1}{{\zeta}_{C_{i,1}}}\\
& \quad {{\bf T}_{1,i}} \succeq 0, {{\bf T}_{1,i}} = {{\bf T}_{1,i}^H}, 1 \le i \le K \\
\end{aligned},
\label{equ:SOA_3}
\end{equation}
where ${\bf A}_{1,i} = {\bf h}_{1,i}^{*}  {\bf h}^T_{1,i}, \in {{\mathcal C}^{N_t \times N_t}}$, ${\bf X}_1 = \sum_{i=1}^{K}{\bf X}_{1,i} \in {{\mathcal C}^{N_t \times L}}$ and ${{\bf T}_{1,i}} = {{\bf X}_{1,i}}{{\bf X}_{1,i}}^H, \in {{\mathcal C}^{N_t \times N_t}}$.
The Lagrange function of ${\mathcal{P}_{\mathsf {MUI}}}$ can be expressed as
\begin{equation}
\begin{aligned}
\mathcal{L} ({\bf X}_1, {u}) & = \frac{N_t}{L} diag([{{\bf T}_{1,1}}, {{\bf T}_{1,2}}, ... ,{{\bf T}_{1,K}}])  \\
& + u \cdot {\zeta}_{C_{i,1}} ({\bf \sigma}_{C_i}^2 + \sum_{j=1,j \ne i}^{K} {\rm tr}( {\bf A}_{1,i} {{\bf T}_{1,j}}) ) \\
&- u \cdot {\rm tr}({\bf A}_{1,i} {{\bf T}_{1,i}} )
\end{aligned}.
\label{equ:SOA_4}
\end{equation}
Letting $\frac{\partial \mathcal{L}}{\partial u} = 0$, we can obtain
\begin{equation}
\begin{aligned}
{\rm tr}({\bf A}_{1,i} {{\bf T}_{1,i}}) =   {\zeta}_{C_{i,1}} ({\bf \sigma}_{C_i}^2 + \sum_{j=1,j \ne i}^{K} {\rm tr}( {\bf A}_{1,i} {{\bf T}_{1,j}}) )
\end{aligned}.
\label{equ:SOA_5}
\end{equation}
And $\frac{\partial \mathcal{L}}{\partial {\bf X}_{1,i}}$ can be expressed as
\begin{equation}
\begin{aligned}
\frac{\partial \mathcal{L}}{\partial {\bf X}_{1,i}} = 2{\bf X}_{1,i}(\frac{N_t}{L} - u \cdot {\bf A}_{1,i}^H )
\end{aligned}.
\label{equ:SOA_6}
\end{equation}
Then, we can obtain the update iteration formula for ${\bf X}_1$ as
\begin{equation}
\begin{aligned}
{\bf X}_{1,i}^{(t)} &= {\bf X}_{1,i}^{(t)} - \kappa_t \frac{\partial \mathcal{L}}{\partial {\bf X}_{1,i}^{(t-1)}} \\
{\bf X}_1^{(t)} &= \sum_{i=1}^{K} {\bf X}_{1,i}^{(t)}
\end{aligned},
\label{equ:SOA_7}
\end{equation}
where $\kappa_t$ is the step length of iteration, which can be obtained by backtracking line search algorithm \cite{[BL]}.

\subsubsection{MBI\_C to users}\label{sec:SOA_2}

Considering MBI\_C from other ISAC base stations, the optimization problem can be expressed as
\begin{equation}
\begin{aligned}
{\mathcal{P}_{\mathsf {MBI\_C}}} : \min_{{{\bf X}_j}} & \quad   \frac{N_t}{L} {diag({{\bf X}_j}{{\bf X}_j}^H)}, j = 2,3  \\
s.t.
& \frac{ {\rm tr}({\bf A}_{2,i} {\bf X}_2 {\bf X}_2^H) + {\rm tr}({\bf A}_{3,i} {\bf X}_3 {\bf X}_3^H)   }   {{\rm tr}({\bf A}_{1,i} {{\bf T}_{1,i}}) } \le \frac{1}{{\zeta}_{C_{i,2}}} \\
& 1 \le i \le K\\
\end{aligned},
\label{equ:SOA_8}
\end{equation}
where ${\bf A}_{2,i} = {\bf h}_{2,i}^{*}  {\bf h}^T_{2,i}, \in {{\mathcal C}^{N_t \times N_t}}$ and ${\bf A}_{3,i} = {\bf h}_{3,i}^{*} {\bf h}^T_{3,i}, \in {{\mathcal C}^{N_t \times N_t}}$.
By solving the Lagrange dual problem for ${\mathcal{P}_{\mathsf {MBI\_C}}}$, we can obtain the update iteration formula for ${\bf X}_2$ and ${\bf X}_3$ as
\begin{equation}
\begin{aligned}
{\bf X}_2^{(t)} &= {\bf X}_2^{(t)} - \kappa_t \frac{\partial \mathcal{L}}{\partial {\bf X}_{2}^{(t-1)}} \\
{\bf X}_3^{(t)} &= {\bf X}_3^{(t)} - \kappa_t \frac{\partial \mathcal{L}}{\partial {\bf X}_{3}^{(t-1)}}
\end{aligned},
\label{equ:SOA_9}
\end{equation}
where
\begin{equation}
\begin{aligned}
\frac{\partial \mathcal{L}}{\partial {\bf X}_{2}^{(t-1)}} &= 2{\bf X}_{2}(\frac{N_t}{L} + u \cdot {\zeta}_{C_{i,2}} {\bf A}_{2,i}^H ) \\
\frac{\partial \mathcal{L}}{\partial {\bf X}_{3}^{(t-1)}} &= 2{\bf X}_{3}(\frac{N_t}{L} + u \cdot {\zeta}_{C_{i,2}} {\bf A}_{3,i}^H ) \\
{\rm tr}({\bf A}_{1,i} {{\bf T}_{1,i}}) &=   {\zeta}_{C_{i,2}} ({\bf \sigma}_{C_i}^2 + {\rm tr}( {\bf A}_{2,i} {\bf X}_2{\bf X}_2^H) \\
&+ {\rm tr}( {\bf A}_{3,i} {\bf X}_3{\bf X}_3^H))
\end{aligned}.
\label{equ:SOA_10}
\end{equation}

\subsubsection{MPI to ISAC $\rm BS_1$}\label{sec:SOA_3}

Considering MPI, the optimization problem can be expressed as
\begin{equation}
\begin{aligned}
{\mathcal{P}_{\mathsf {MPI}}} : \min_{{{\bf X}_1}} & \quad  \frac{1}{L}  {diag({{\bf X}_1}{{\bf X}_1}^H)}  \\
s.t.
&
\frac  {  \sum_{l=1}^{L_{p}-1}  {\rm tr}({\bf B}_{1,l} {{\bf T}_{1}} ) + {\bf \sigma}_{R_1}^2  } { {\rm tr}({\bf B}_{1,0} {{\bf T}_{1}} )} \le \frac{1}{{\zeta}_{R_{1,1}}}
\\
\end{aligned}.
\label{equ:SOA_11}
\end{equation}
For the MPI constraint in ${\mathcal{P}_{\mathsf {MPI}}}$, we define its dual variables as $u \ge 0$. Then, the Lagrange function of ${\mathcal{P}_{\mathsf {MPI}}}$ can be expressed as
\begin{equation}
\begin{aligned}
\mathcal{L} ({\bf X}_1, {u}) & = \frac{N_t}{L} diag({\bf X}_1 {\bf X}_1^H) \\
& - u \cdot {\rm tr}({\bf B}_{1,0} {{\bf X}_1} {{\bf X}_1^H})\\
& + u  {\zeta}_{R_{1,1}} ({\bf \sigma}_{R_1}^2 + \sum_{l=1}^{L_{p}-1} {\rm tr}( {\bf B}_{1,l} {{\bf X}_1} {{\bf X}_1^H}) )
\end{aligned},
\label{equ:SOA_12}
\end{equation}
where ${\bf B}_{1,l} = {{\bf G}_{1,1,l}^{*}}  {{\bf G}{_{1,1,l}^T}}, \in {{\mathcal C}^{N_t \times N_t}} $ and ${\bf B}_{1,0} = {{\bf G}_{1,1,0}^{*}}  {{\bf G}{_{1,1,0}^T}} ,\in {{\mathcal C}^{N_t \times N_t}} $.
Letting $\frac{\partial \mathcal{L}}{\partial u} = 0$, we can obtain
\begin{equation}
\begin{aligned}
{\rm tr}({\bf B}_{1,0} {{\bf X}_1} {{\bf X}_1^H}) =   {\zeta}_{R_{1,1}} ({\bf \sigma}_{R_1}^2 + \sum_{l=1}^{L_{p}-1} {\rm tr}( {\bf B}_{1,l} {{\bf X}_1} {{\bf X}_1^H}) )
\end{aligned}.
\label{equ:SOA_13}
\end{equation}
And $\frac{\partial \mathcal{L}}{\partial {\bf X}_1}$ can be expressed as
\begin{equation}
\begin{aligned}
\frac{\partial \mathcal{L}}{\partial {\bf X}_1} = 2{\bf X}_1(\frac{N_t}{L} + u \cdot (-{\bf B}_{1,0}^H + {\zeta}_{R_{1,1}} \sum_{l=1}^{L_{p}-1} {\bf B}_{1,l}^H  ))
\end{aligned}.
\label{equ:SOA_14}
\end{equation}
Then, we can obtain the update iteration formula for ${\bf X}_1$ as
\begin{equation}
\begin{aligned}
{\bf X}_1^{(t)} = {\bf X}_1^{(t)} - \kappa_t \frac{\partial \mathcal{L}}{\partial {\bf X}_1^{(t-1)}}
\end{aligned},
\label{equ:SOA_15}
\end{equation}
where $\kappa_t$ is the step length of iteration.
\subsubsection{MBI\_R to ISAC $\rm BS_1$}\label{sec:SOA_4}

Considering MBI\_R from other ISAC base stations, the optimization problem can be expressed as
\begin{equation}
\begin{aligned}
{\mathcal{P}_{\mathsf {MBI\_R}}} : \min_{{{\bf X}_2}} & \quad  \frac{N_t}{L} {diag({{\bf X}_j}{{\bf X}_j}^H)}, j = 2,3   \\
s.t.
& \frac  { {\rm tr}({\bf B}_{2} {{\bf T}_{2}} ) + {\rm tr}({\bf B}_{3} {{\bf T}_{3}} )  } { {\rm tr}({\bf B}_{1,0} {{\bf T}_{1}} )} \le \frac{1}{{\zeta}_{R_{1,2}}}\\
\end{aligned},
\label{equ:SOA_16}
\end{equation}
where ${\bf B}_{2} = {{\bf G}_{1,2}^{*}}  {{\bf G}{_{1,2}^T}}, \in {{\mathcal C}^{N_t \times N_t}} $ and ${\bf B}_{3} = {{\bf G}_{1,3}^{*}}  {{\bf G}{_{1,3}^T}}, \in {{\mathcal C}^{N_t \times N_t}} $.
By solving the Lagrange dual problem for ${\mathcal{P}_{\mathsf {MBI\_R}}}$, we can obtain the update iteration formula for ${\bf X}_2$ and ${\bf X}_3$ as
\begin{equation}
\begin{aligned}
{\bf X}_2^{(t)} &= {\bf X}_2^{(t)} - \kappa_t \frac{\partial \mathcal{L}}{\partial {\bf X}_{2}^{(t-1)}} \\
{\bf X}_3^{(t)} &= {\bf X}_3^{(t)} - \kappa_t \frac{\partial \mathcal{L}}{\partial {\bf X}_{3}^{(t-1)}}
\end{aligned},
\label{equ:SOA_17}
\end{equation}
where
\begin{equation}
\begin{aligned}
\frac{\partial \mathcal{L}}{\partial {\bf X}_{2}^{(t-1)}} &= 2{\bf X}_{2}(\frac{N_t}{L} + u \cdot {\zeta}_{R_{1,2}} {\bf B}_{2,0}^H ) \\
\frac{\partial \mathcal{L}}{\partial {\bf X}_{3}^{(t-1)}} &= 2{\bf X}_{3}(\frac{N_t}{L} + u \cdot {\zeta}_{R_{1,2}} {\bf B}_{3,0}^H ) \\
{\rm tr}({\bf B}_{1,0} {\bf X}_1{\bf X}_1^H) &=   {\zeta}_{R_{1,2}} ({\bf \sigma}_{R_1}^2 + {\rm tr}( {\bf B}_{2} {\bf X}_2{\bf X}_2^H) \\ &+ {\rm tr}( {\bf B}_{3} {\bf X}_3{\bf X}_3^H) )
\end{aligned}.
\label{equ:SOA_18}
\end{equation}
The threshold values satisfy the following relationship
\begin{equation}
	\begin{aligned}
		\frac{1}{{\zeta}_{C_i}} &= \frac{1}{{\zeta}_{C_{i,1}}} + \frac{1}{{\zeta}_{C_{i,2}}} \\
		\frac{1}{{\zeta}_{R_1}} &= \frac{1}{{\zeta}_{R_{1,1}}} + \frac{1}{{\zeta}_{R_{1,2}}}
	\end{aligned}.
	\label{equ:SOA_20}
\end{equation}
%
The proposed optimization algorithm SOA can be summarized by the following Algorithm \ref{alg:SOA_1}.
\begin{algorithm}[htb]
	\caption{Sequential optimization algorithm (SOA)}
	\label{alg:SOA_1}
	\begin{algorithmic}
		\Require
		$\bf H$, ${\zeta }_{C,i}$, ${\zeta }_{R_j}$ and $N_t$;
		\State \quad 1). According to the relationship between communication and radar sensing channel parameters listed in table \ref{label:ISAC_CSI}, generate $\bf G$ by \eqref{equ:Rad_CSI_1} and \eqref{equ:Rad_CSI_2};
		\State \quad 2). Rewrite the reciprocal of SINR by \eqref{equ:SOA_1} and \eqref{equ:SOA_2};
		\State Initialization: Set the iteration threshold to be $T = 100$ and the number of iterations $t=1$;
		\While{$ t \le T$}
		\State  a). Update ${\bf X}_{1}^{(t)}$ based on \eqref{equ:SOA_5}, \eqref{equ:SOA_6} and \eqref{equ:SOA_7};
		\State  b). Update ${\bf X}_{2}^{(t)}$ and ${\bf X}_{3}^{(t)}$ based on \eqref{equ:SOA_9} and \eqref{equ:SOA_10} after substituting ${\bf X}_{1}^{(t)}$ into ${\mathcal{P}_{\mathsf {MBI\_C}}}$;
		\State c). Update ${\bf X}_{1}^{(t)}$ based on \eqref{equ:SOA_13} and \eqref{equ:SOA_15};
		\State d). Update ${\bf X}_{2}^{(t)}$ and ${\bf X}_{3}^{(t)}$ based on \eqref{equ:SOA_17} and \eqref{equ:SOA_18} after substituting ${\bf X}_{1}^{(t)}$ into ${\mathcal{P}_{\mathsf {MBI\_R}}}$;
		\If {$\gamma_{R_j} \ge \zeta_{R_j}$ and $\gamma_{C,i} \ge \zeta_{C,i}, \forall i$}
		 Break.
		\EndIf
		\State $t = t + 1$.
		\EndWhile
		\Ensure ${\bf X}_j, j = 1,2,3$.
	\end{algorithmic}
\end{algorithm}
SOA for solving the other 5 problems is similar to JOA for solving the other 5 problems, as listed in Table \ref{alg:six_JOA}.

\subsection{Convergence Property and Complexity Analysis}\label{sec:CA}
We note here that JOA is a type of the semidefinite relaxation (SDR) algorithm \cite{[Liu_1]}, which can be solved by MATLAB's cvx tool.
SOA is a feasible descent algorithm for the Lagrange dual problem of the collaborative precoding problem {\color{black} ${\mathcal{P}^{\mathsf {DI}}_{\mathsf {PPC}}}$}. We can transform {\color{black} ${\mathcal{P}^{\mathsf {DI}}_{\mathsf {PPC}}}$} into a convex optimization problem by omitting the rank-1 constraint, and obtain an approximate solution by eigenvalue decomposition. Therefore, SOA also meets the convergence condition, and our simulation results show that SOA converges to a near-optimal point with less than 20 iterations at a modest accuracy.

The computational complexity of JOA and SOA is constrained by two aspects. One is the threshold constraint of SINR, the other is the constraint of the number of users $3K$ and the number of antennas $N_t$. The former mainly affects the number of iterations of SDP optimization problem, while the latter mainly affects the complexity of matrix operation in each iteration process. Since there is no closed-form expression for the iterative complexity of SDP optimization, we use the average execution time to reflect it.
For the complexity of matrix operation in each iteration optimization process, we analyze it in terms of flops, which is defined as multiplication, division, or addition of two floating point numbers \cite{[Flop_43]}.
According to the definition in \cite{[Flop_43]}, a complex addition and a complex multiplication can be considered as 2 and 6 flops, respectively. An addition between two $M × N$ complex matrices consists of $2MN$ flops, a multiplication between $M × N$ and $N × L$ complex matrices requires $8MNL$ operations.
In all calculations, we consider only the highest order terms and omit the lower order terms.
Then, the complexity of JOA and SOA for 6 optimization problems can be listed in Table \ref{Parameter:complexity_2}.
Since the SDR algorithm does not have a complexity representation in closed form, we cannot analytically compare the two series. Therefore, a simulation-based comparison is presented in section \ref{sec:Simulation_3}, which highlights the complexity savings of SOA.

\begin{table}[h]
	\caption{Computational complexity of JOA and SOA for 6 optimization problems.}
    \centering
	\label{Parameter:complexity_2}
	\begin{tabular}{c|c|c}
		\hline
		\hline
		 \qquad Problem \qquad \qquad & JOA $\mathcal{O} (\cdot)$ & \qquad SOA $\mathcal{O} (\cdot)$  \qquad \qquad \\ [1.1ex] \hline
		{\color{black} ${\mathcal{P}^{\mathsf {DI}}_{\mathsf {TPC}}}$} & $3N_t^2K + 9N_tK^2$ &  $3N_t^2K$ \\ [1.1ex] \hline
		{\color{black} ${\mathcal{P}^{\mathsf {DI}}_{\mathsf {PPC}}}$} & $3N_t^2K + 9N_tK^2$ &  $3N_t^2K$ \\ [1.1ex] \hline
		{\color{black} ${\mathcal{P}^{\mathsf {DI}}_{\mathsf {CMC}}}$} & $3N_t^2K + 9N_tK^2 + 2^{N_t}$ & $3N_t^2K + 2^{N_t}$ \\ [1.2ex] \hline
		{\color{black} ${\mathcal{P}^{\mathsf {CI}}_{\mathsf {TPC}}}$} & $3N_t^2K + 9N_tK^2$ &  $3N_t^2K$ \\ [1.1ex] \hline
		{\color{black} ${\mathcal{P}^{\mathsf {CI}}_{\mathsf {PPC}}}$} & $3N_t^2K + 9N_tK^2$ &  $3N_t^2K$ \\ [1.1ex] \hline
		{\color{black} ${\mathcal{P}^{\mathsf {CI}}_{\mathsf {CMC}}}$} & $3N_t^2K + 9N_tK^2 + 2^{N_t}$ & $3N_t^2K + 2^{N_t}$  \\ [1.1ex] \hline
	\end{tabular}
\end{table}

\section{Simulation Results}\label{sec:Simulation}

In this section, we will first analyze the communication and sensing performance of the collaborative precoding design under different constraints and then analyze the trade-off between the communication and sensing performance. In the end, we will compare the complexity of JOA and SOA algorithms.
Simulation parameters used in this section are shown in table \ref{Parameter:simulation} \cite{[Liu_parameters_1],[Strum_parameters_1]}.

\renewcommand{\arraystretch}{1.1}
\begin{table}[t]
	\caption{Simulation parameters adopted in this paper.}
    \centering
	\label{Parameter:simulation}
	\begin{tabular}{l|l|l}
		\hline
		\hline
		Items & Value & Meaning of the parameter \\ \hline
		$f_c$ & $24$ GHz \cite{[Strum_parameters_1]} & Frequency of ISAC signals \\ \hline
		${\bf X}_0$ & Orthogonal
		chirp \cite{[Liu_1],[OCDM]} & Reference radar signal \\ \hline
		$B$ & 100 MHz \cite{[Strum_parameters_1]} & System bandwidth \\ \hline
		${\bf \sigma}_R$ & -94 dBm & Thermal noise of radar channel \\ \hline
		$N_t$ & $10\sim40$ \cite{[Liu_parameters_1]}& Number of antenna array elements \\ \hline
		${\bf \sigma}_C$ & -94 dBm & Noise of communication channel \\ \hline
		$K$ & 5 \cite{[Liu_parameters_1]}& Number of users  \\ \hline
		${\zeta}_{R_j}$ & $0\sim30$ dB & Threshold of radar SINR \\ \hline
		$L$ & 50 \cite{[Liu_parameters_1]}& Length of data stream \\ \hline ${\zeta}_{C_i}$ & $0\sim30$ dB & Threshold of communication SINR \\ \hline
		$L_p$ & 3 & Number of interference sources \\ \hline
		$\theta_k$ & $10^o\sim20^o$ & Angle of users relative to BS \\ \hline $\epsilon$ & $0 \sim 2$ \cite{[SC_37]}& Threshold of CMC and SC \\ \hline
		$T$ & $100$ & Iteration threshold \\ \hline
		$\kappa_t$ & $0.01$ & Step length of iteration \\ \hline
	\end{tabular}
\end{table}

\subsection{Communication and Sensing Performance}\label{sec:Simulation_1}

\begin{figure}[h]
	\includegraphics[scale=0.6]{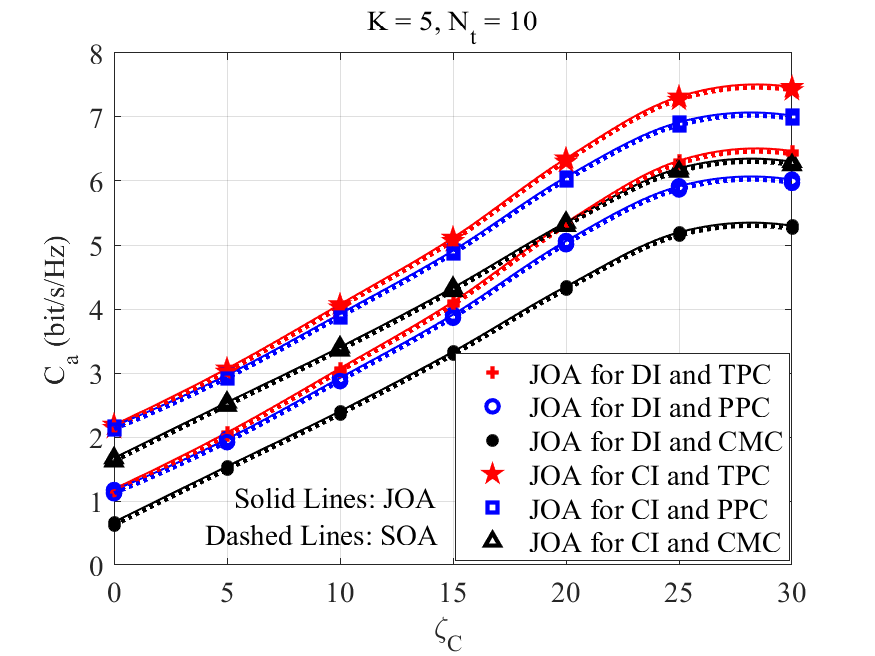}
	\centering
	\caption{Average communication rate for different ${\zeta}_C$ with ${\zeta}_R=10$ dB.}
	\label{fig:C_1}
\end{figure}

\begin{figure}[h]
	\includegraphics[scale=0.6]{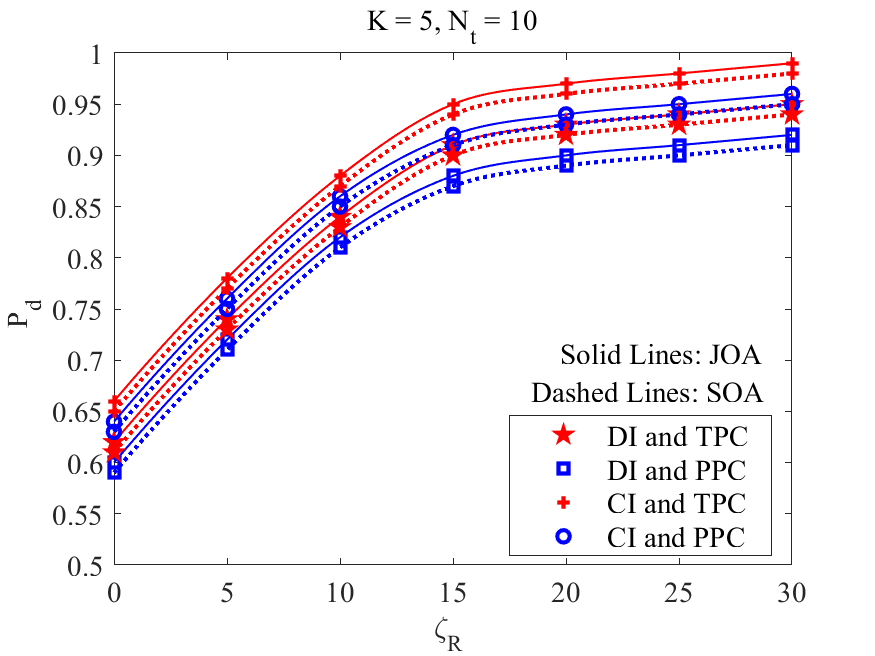}
	\centering
	\caption{Detection probability for different ${\zeta}_R$ with ${\zeta}_C=10$ dB.}
	\label{fig:PD_1}
\end{figure}

\begin{figure}[h]
	\includegraphics[scale=0.6]{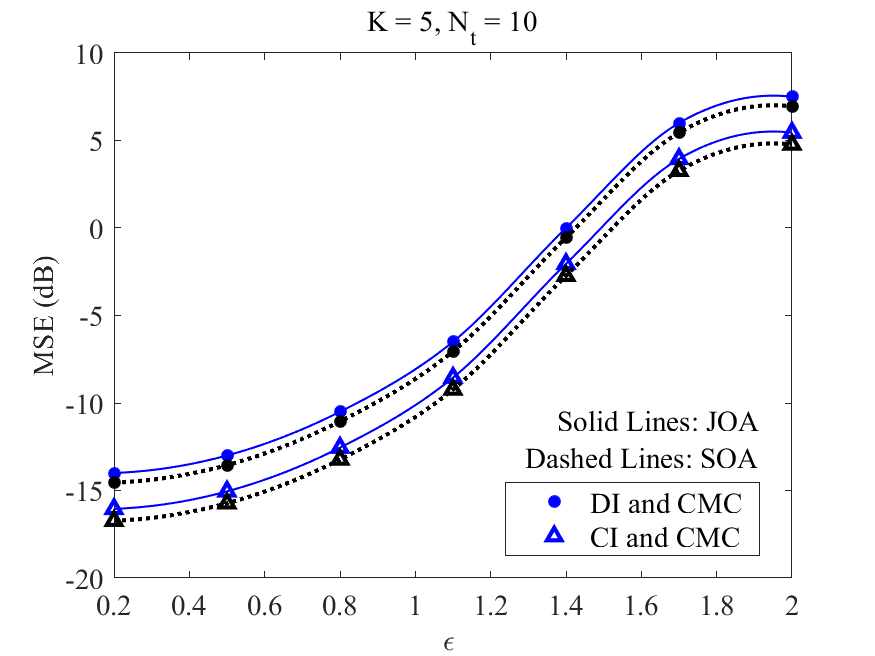}
	\centering
	\caption{MSE of optimized waveform and reference waveform for different $\epsilon$ with ${\zeta}_R=10$ dB and ${\zeta}_C=10$ dB.}
	\label{fig:MSE_1}
\end{figure}

\begin{figure}[htbp]
	\centering
	\begin{minipage}[t]{0.48\textwidth}
		\centering
		\includegraphics[width=7.5cm]{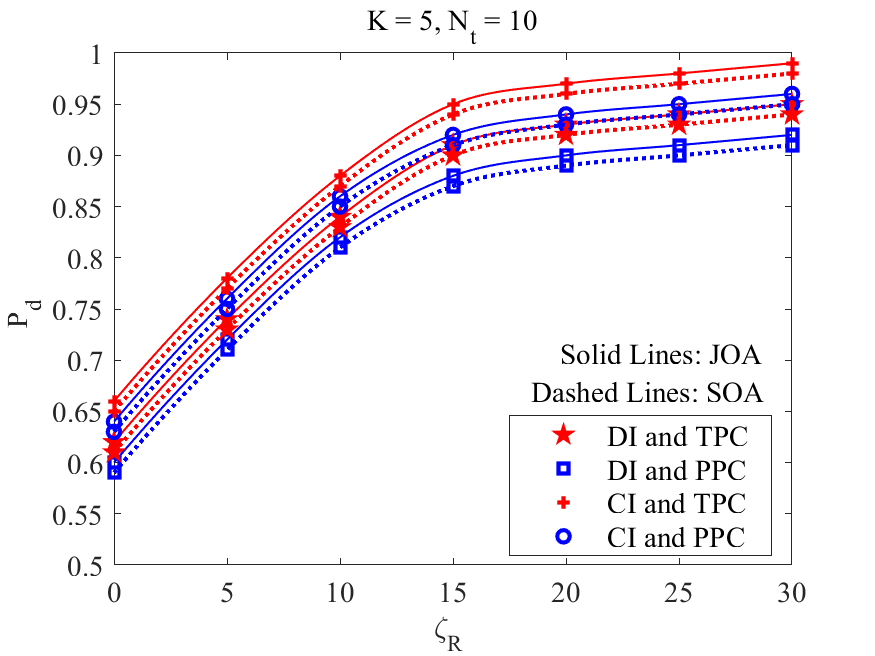}
		\caption{Detection probability for different ${\zeta}_R$ with ${\zeta}_C=10$ dB.}
		\label{fig:PD_1}
	\end{minipage}
	\begin{minipage}[t]{0.48\textwidth}
		\centering
		\includegraphics[width=7.5cm]{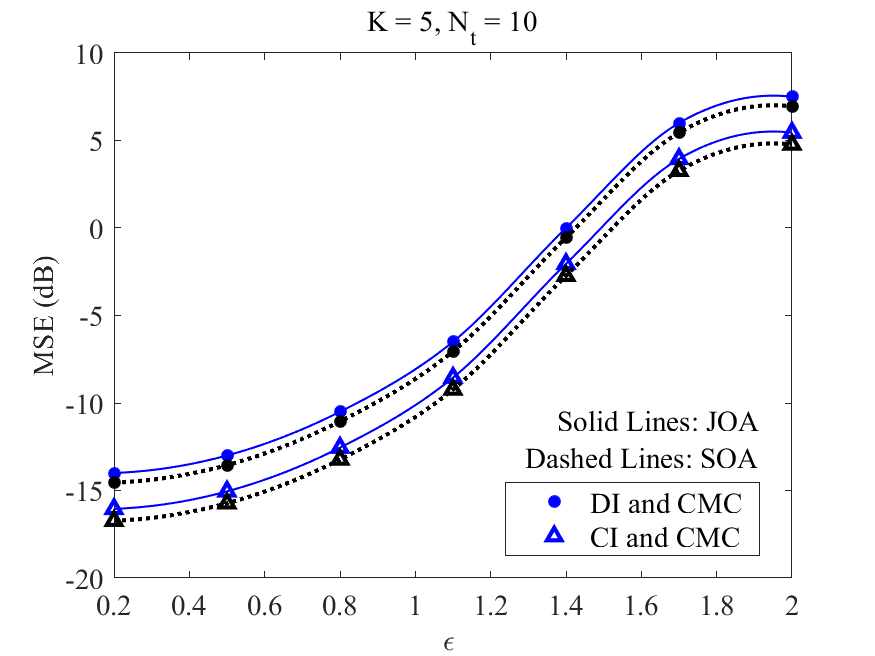}
		\caption{MSE of optimized waveform and reference waveform for different $\epsilon$ with ${\zeta}_R=10$ dB and ${\zeta}_C=10$ dB.}
		\label{fig:MSE_1}
	\end{minipage}
\end{figure}

%
%

\subsubsection{Average communication rate}\label{sec:Simulation_1_1}

For the sake of presentation, we assume that the communication SINR threshold of all users is the same, i.e. ${\zeta}_{C} = {\zeta}_{C,i}, \forall i$.
We first show the DL communication performance obtained by
different approaches in Fig. \ref{fig:C_1} in term of the average communication rate ${C}_a$, which can be defined as
\begin{equation}
\begin{aligned}
{C}_a = \frac{\sum_{i=1}^{3K}{log_2 (1+\gamma_{C_i})} }{3K}
\end{aligned}.
\label{equ:Sim_1}
\end{equation}
Fig. \ref{fig:C_1} shows the average communication rate for different ${\zeta}_C$ with $K=5$, $N_t=10$ and ${\zeta}_R=10$ dB.
Following the simulation configurations in \cite{[Liu_1],[Liu_38]}, we employ the orthogonal
chirp waveform matrix as the reference signal.
It should be noted that the orthogonal chirp signal is one of ISAC signals and has the ability to transmit communication \cite{[ISAC_chrip]}. The average communication rate under CMC can also expressed approximatively by \eqref{equ:Sim_1}, which can be verified in figure 9 of \cite{[Liu_1]}.

It can be observed in Fig. \ref{fig:C_1} that the average communication rate is increasing with increasing ${\zeta}_C$, which means that both JOA and SOA can significantly improve the average communication rate compared to the situation without collaborative precoding.
In terms of the same type of curves, the dashed line is close to the solid line, which means that the performance of SOA is close to JOA.
Moreover, the collaborative precoding design based on CI constraint can obtain a relatively higher average communication rate than that for DI, which is consistent with the theoretical analysis.

\subsubsection{Detection probability}\label{sec:Simulation_1_2}
For collaborative precoding design based on TPC and PPC, the detection probability $P_d$ is used as the metric, where we consider the constant false-alarm rate detection for point-like targets from $L_p-1$ path. The false alarm probability for ISAC base station is $P_f = 10^{-7}$. And we can obtain the detection probability $P_d$ based on \cite{[PD_9]} eq. (69).
Fig. \ref{fig:PD_1} shows the detection probability for different ${\zeta}_R$ with $K=5$, $N=10$ and ${\zeta}_C=10$ dB.
It can be observed in Fig. \ref{fig:PD_1} that the detection probability is increasing with increasing ${\zeta}_R$, which means that both JOA and SOA can significantly improve the detection probability compared to the situation without collaborative precoding.

\subsubsection{MSE of optimized waveform and reference waveform}\label{sec:Simulation_1_3}
For collaborative precoding design based on CMC, the mean square error (MSE) of the optimized ISAC signal and reference radar signal is used as the metric,
which can be defined as
\begin{equation}
	\begin{aligned}
		{\rm MSE} = \frac{\sum_{j=1}^{3}  {|| {\bf X}_j - {\bf X}_0    ||^2 }}{3}
	\end{aligned}.
	\label{equ:MSE}
\end{equation}
Fig. \ref{fig:MSE_1} shows MSE for different $\epsilon$ with $K=5$, $N=10$, ${\zeta}_R=10$ dB and ${\zeta}_C=10$ dB. With the increase of constraint threshold $\epsilon$, the restriction on waveform becomes weaker, MSE also increases, and the sensing performance decreases.

\subsection{Trade-off between Communication Performance and Sensing Performance}\label{sec:Simulation_2}

In order to analyze the tradeoff between collaborative precoding design in communication and sensing performance, we furthermore investigate the trade-off between average communication rate and detection probability, and the trade-off between average communication rate and MSE of the optimized ISAC signal and reference radar signal.

\begin{figure}[h]
	\includegraphics[scale=0.6]{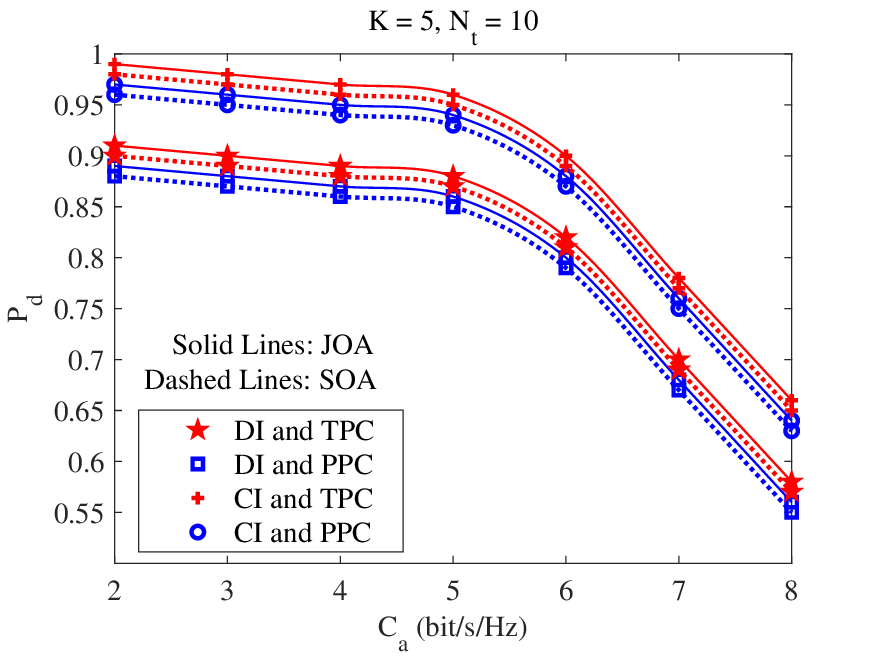}
	\centering
	\caption{Trade-off between communication rate and detection probability.}
	\label{fig:C_PD_1}
\end{figure}

\begin{figure}[h]
	\includegraphics[scale=0.6]{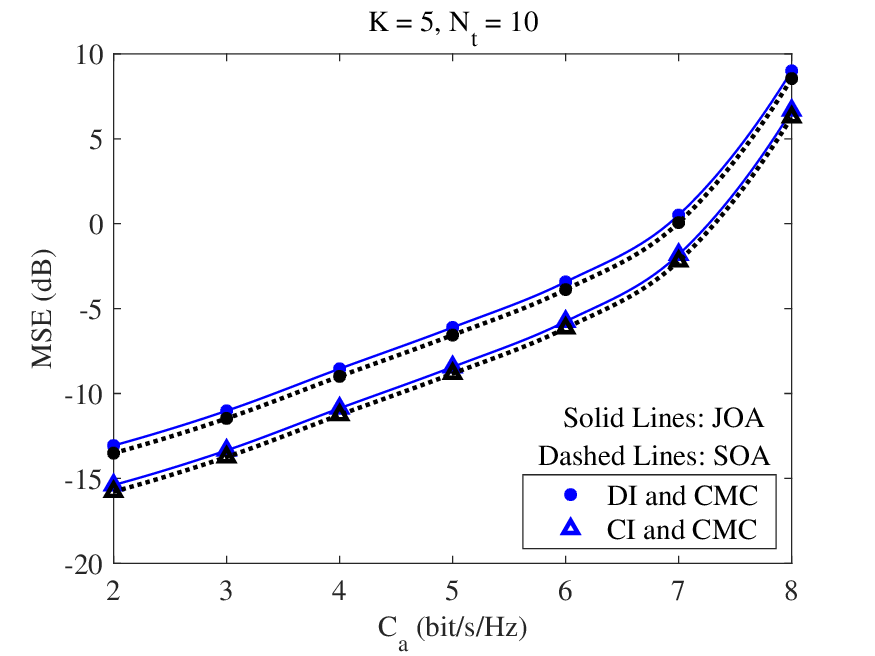}
	\centering
	\caption{Trade-off between communication rate and MSE.}
	\label{fig:C_MSE_1}
\end{figure}

As Fig. \ref{fig:C_PD_1} shows that $P_d$ is decreasing with the increasing $C_a$, which means that there exists a trade-off between the communication rate and the radar detection performance.
The same trends appear in Fig. \ref{fig:C_MSE_1}, where we employ the
MSE of the optimized ISAC signal and reference radar signal as the radar metric. Both Fig. \ref{fig:C_PD_1} and Fig. \ref{fig:C_MSE_1} prove that our approach can achieve a favorable trade-off between radar and communication.

%

\subsection{Comparison of the  Complexity of JOA and SOA}\label{sec:Simulation_3}

\begin{figure}[h]
	\includegraphics[scale=0.6]{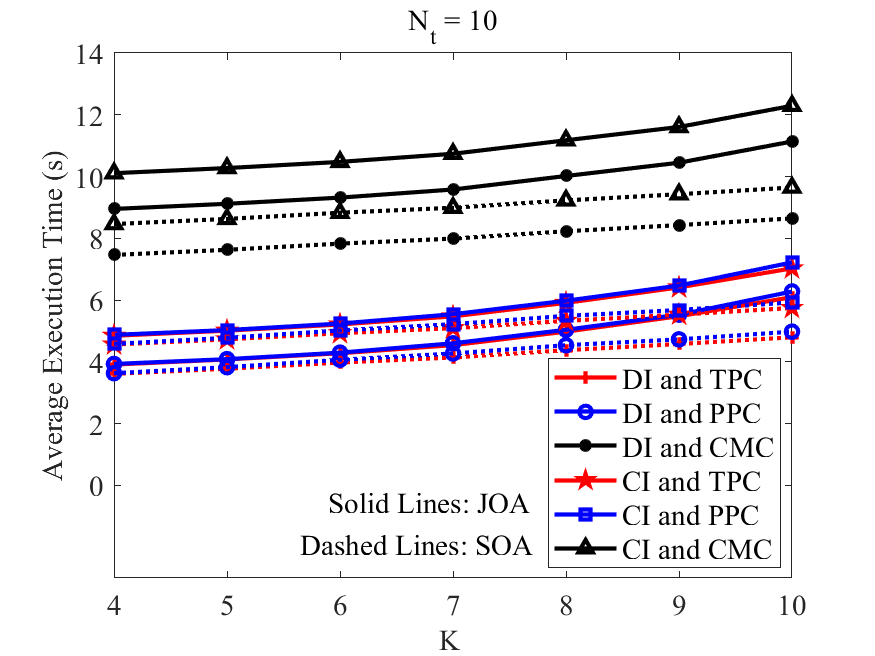}
	\centering
	\caption{Average execution time for different $K$ with $N_t=10$, ${\zeta}_R=10$ dB and ${\zeta}_C=10$ dB.}
	\label{fig:Complexity_K_1}
\end{figure}

\begin{figure}[h]
	\includegraphics[scale=0.6]{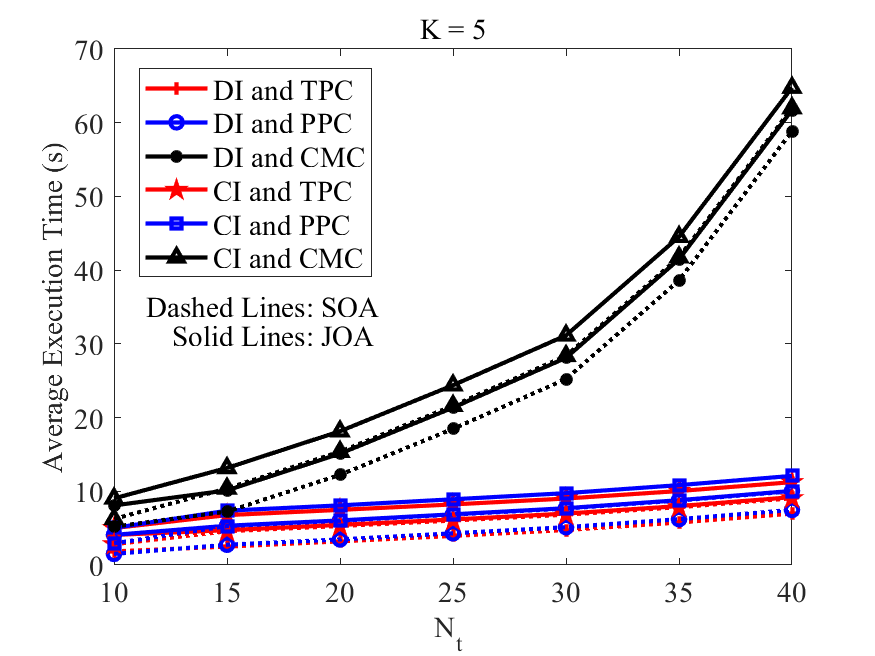}
	\centering
	\caption{Average execution time for different $N_t$ with $K=5$, ${\zeta}_R=10$ dB and ${\zeta}_C=10$ dB.}
	\label{fig:Complexity_N_1}
\end{figure}

In section \ref{sec:CA}, we have briefly compared the complexity of the collaborative precoding design of JOA and SOA algorithms under different constraints from the perspective of flops. We will furthermore compare complexity performance of the two algorithms by average execution time. It should be pointed out that due to the error of computer running time, the average execution time for each case is the average of multiple tests.
In Fig. \ref{fig:Complexity_K_1}, we explore the average execution time of both algorithms for different users with the number of antenna array elements $N=10$, threshold of radar SINR ${\zeta}_R=10$ dB and communication SINR ${\zeta}_C=10$ dB.
As Fig. \ref{fig:Complexity_K_1} shows, we can get the following conclusions:

$\cdot$ With the number of users $K$ increases, the average execution time of both algorithms increases because of the increasing of the number of users SINR constraints, which is consistent with the theoretical analysis.

$\cdot$ Using the same algorithm, it takes less time to solve the precoding design optimization problem with TPC constraint than that with PPC constraint, because the latter need to divide the diagonal constraint of TPC into N quadratic equality constraints. It takes much more time to solve the precoding design optimization problem with CMC constraint, because the former needs to solve the NP-hard problem based on BnB methodology, which is of high complexity. Moreover, it takes more time to solve the precoding design optimization problem with CI than without that, because the latter has an extra step of real representation.

$\cdot$ Using different algorithms, it takes less time to adopt SOA than JOA for precoding design optimization under the same constraints, which verifies the advantage of SOA in algorithm complexity.

In Fig. \ref{fig:Complexity_N_1}, we furthermore explore the average execution time of both algorithms for different the number of antenna array elements with $K=5$, threshold of radar SINR ${\zeta}_R=10$ dB and communication SINR ${\zeta}_C=10$ dB. We find that as the number of antenna array elements $N$ increases, the average execution time of solving the precoding design optimization under CMC constraint is increasing much faster than that of DI and CI constraints.


\subsection{Comparison of SOA and other typical algorithms}\label{sec:Simulation_4}
{\color{black}
We mainly compare the proposed JOA with the existing benchmark schemes from two perspectives:
\begin{itemize}
	\item On the one hand, most of the existing collabortive precoding algorithms are for communication systems, there are relatively few benchmark collabortive precoding schemes designed for mutual interference between ISAC base stations.
	Moreover, the JOA proposed in this paper is similar to typical collaborative precoding algorithms used in communication systems, except that the problems addressed are different.
	Therefore, JOA can be regarded as a benchmark collaborative precoding algorithm to solve the mutual interference problem among ISAC base stations.
	Then, many simulation results comparing the sensing and communication performance of SOA and JOA have been provided in this paper, such as Fig. 7 and Fig. 8.
	\item On the other hand, there are some  existing collabortive precoding algorithms for single ISAC BS \cite{[Liu_1]}. We simulate and compare the sensing and communication performance of SOA and existing typical collaborative precoding algorithms.
	The number of ISAC base station is set as 1, the number of antenna array elements is set as 16, the number of users is set as 6.
	As Fig. \ref{fig:compare} shows, for the single ISAC base station scenario, the joint communication and sensing performance obtained by JOA is close to the precoding algorithm proposed in \cite{[Liu_1]}, and slightly higher than SOA.
\end{itemize}

\begin{figure}[h]
	\includegraphics[scale=0.6]{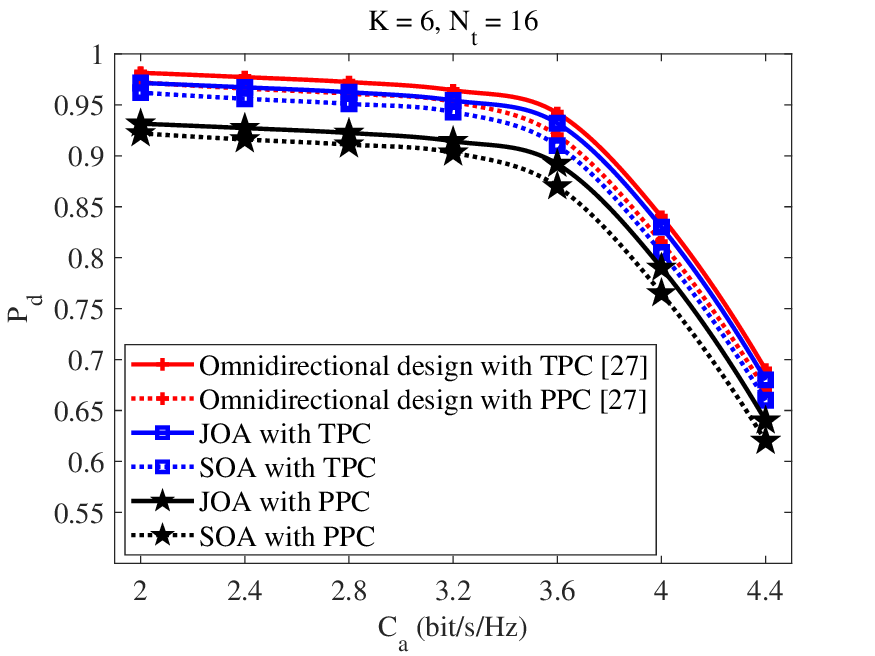}
	\centering
	\caption{Trade-off between communication rate and detection probability with different algorithms.}
	\label{fig:compare}
\end{figure}

}
%

\section{Conclusion and Discussion}\label{sec:Conclusion}

In this paper, we establish a mutual interference model for adjacent ISAC base stations that can communicate with cellular users and detects radar targets simultaneously.
The mutual interference model consists of DL communication interference and radar sensing interference. DL communication interference includes multiuser interference and inter-base station DL communication interference. Radar sensing interference includes multipath interference and echo interference among ISAC base stations.
To design the precoding for ISAC base station, we propose two optimization algorithms, JOA and SOA. JOA directly solves the optimization problem after transforming the NP-hard problem into convex optimization problem. While SOA divides the precoding design optimization problem into four subproblems first and then obtain the optimal solution by gradient descent. We compare and analyze the computational complexity of JOA and SOA. Simulation results confirm that SOA can reduce the computational complexity of the algorithm to a certain extent.
Moreover, we evaluate the proposed collaborative precoding design approaches by considering sensing and DL communication performance via numerical results.

{\color {black}
It should be noted that since ISAC base stations use continuous wave signals for sensing, which requires the ability to transmit ISAC signals and receive echoes simultaneously. Then, the transmitting ISAC signal may leak directly into the receiving antenna, which in turn generates self-interference.
Self-interference is indeed a typical problem in radar sensing interference, and there has been related work to study the problem in depth.
\begin{itemize}
	\item Physical isolation: Install electromagnetic shielding devices between the transmitting antenna array and the receiving antenna array to reduce self-interference \cite{[Xu]}. At present, this program is still in the research stage, which is difficult to perfectly solve the problem of self-interference.
	\item Interference cancellation: The receiver can adopt interference cancellation algorithms to eliminate the transmit signal leaked from the transmitter \cite{[SIC]}. At present, the high complexity of this scheme is an important factor limiting its application.
	\item Precoding design: With the optimization objective of minimizing self-interference, the reasonable precoding design is carried out to mitigate the impact of self-interference on the sensing performance. At present, this field is still in the initial stage of research, and some solutions have been proposed \cite{[SI-1],[SI-2]}.
\end{itemize}
In this paper, we focus on the mutual interference among ISAC base stations and do not conduct an in-depth study on the self-interference of ISAC base stations.
We will further consider the effect of self-interference based on this paper and improve the interference channel model and collaborative precoding algorithm, mainly from the following two aspects:
\begin{itemize}
	\item In terms of the channel model, we need to supplement the self-interference channel from the transmitter side of ISAC base station directly to the receiver side, and then update the SINR of the target echo received by ISAC base station.
	\item In terms of precoding design, the JOA algorithm needs to be supplemented to consider the self-interference constraints, and the complexity will be enhanced; the SOA algorithm needs to consider the self-interference sub-optimization problem and increase the iterative optimization steps.
\end{itemize}
}


%

\bibliographystyle{IEEEtran}
\bibliography{reference}


\ifCLASSOPTIONcaptionsoff
  \newpage
\fi

\end{document}